\documentclass{aa}  

\usepackage{graphicx}
\usepackage{txfonts}
\usepackage{siunitx}
\usepackage{hyperref}
\usepackage{tikz}

\usepackage{xcolor}

\newcommand{\U}[1]{\textcolor{black}{#1}}
\usepackage{subfigure}
\begin{document}

   \title{MEGS: Morphological Evaluation of Galactic Structure}

   \subtitle{Principal Component Analysis as a galaxy morphology model}

   \author{U. Çakır
          \inst{1,2}
          \and
          T. Buck\inst{1,2}
          }

   \institute{Interdisciplinary Center for Scientific Computing (IWR), University of Heidelberg,
 Im Neuenheimer Feld 205, D-69120 Heidelberg\\
 \email{ufuk.cakir@stud.uni-heidelberg.de}
 \and
 Universität Heidelberg, Zentrum für Astronomie, Institut für Theoretische Astrophysik, Albert-Ueberle-Straße 2, D-69120 Heidelberg, Germany\\
 \email{tobias.buck@iwr.uni-heidelberg.de}
             }

   \date{Received Month, XXXX; accepted Month Day, XXXX}

% \abstract{}{}{}{}{} 
% 5 {} token are mandatory
 \abstract
   {Understanding the morphology of galaxies is a critical aspect of astrophysics research, providing insight into the formation, evolution, and physical properties of these vast cosmic structures. Various observational and computational methods have been developed to quantify galaxy morphology, and with the advent of large galaxy simulations, the need for automated and effective classification methods has become increasingly important.}
   {This paper investigates the use of Principal Component Analysis (PCA) as an interpretable dimensionality reduction algorithm for galaxy morphology using the IllustrisTNG cosmological simulation dataset with the aim of developing a generative model for galaxies.}
   {We first generate a dataset of 2D images and 3D cubes of galaxies from the IllustrisTNG simulation, focusing on the mass, metallicity, and stellar age distribution of each galaxy. PCA is then applied to this data, transforming it into a lower-dimensional image space, where closeness of data points corresponds to morphological similarity.}
   {We find that PCA can effectively capture the key morphological features of galaxies, with a significant proportion of the variance in the data being explained by a small number of components. With our method we achieve a dimensionality reduction by a factor of $\sim200$ for 2D images and $\sim3650$ for 3D cubes at a reconstruction accuracy below five percent. }
   {Our results illustrate the potential of PCA in compressing large cosmological simulations into an interpretable generative model for galaxies that can easily be used in various downstreaming tasks such as galaxy classification and analysis.}

\maketitle
%-------------------------------------------------------------------
\section{Introduction}
Galaxy morphology, the study of the structural characteristics and visual appearance of galaxies, has been a significant research topic in astrophysics for many years. Early classification schemes, such as the Hubble sequence \citep{Hubble_1926}, provided a visual framework for classifying galaxies based on their visual features. Historically, galaxy morphology has also been used to identify different structural components of a galaxy through an analysis of its light profile. This process is known as structural decomposition and was first presented in an analysis by \cite{Vaucouler1958}. However, these classification methods often struggled to capture the complexity and diversity of galaxy morphologies and were unable to scale with the increasing size and quality of astronomical datasets.

Fundamentally, galaxy morphology is the phenomenological realization of the underlying fundamental differences in physical properties. The visual appearance of galaxies is heavily influenced by the joint distribution of stellar mass, metallicity, and age due to the nonlinear dependence of stellar luminosity on these fundamental parameters \citep[e.g.][]{Kuiper1938}. This further implies that different galaxy components trace different formation episodes of galaxies. In general, galaxy light can be modeled parametrically as either a single component (usually with a \citealt{Sersic1963} profile), or with two or more components \citep[e.g.][]{Cook2019}. However, in the current era of large-scale cosmological simulations, a vast amount of data are available, allowing us to investigate galaxies and their appearance in exceptional detail. 

\begin{figure*}
   %\sidecaption
   \subfigure[2D]{
       \includegraphics[width=0.49\hsize, trim={0 0 0 0cm},clip]{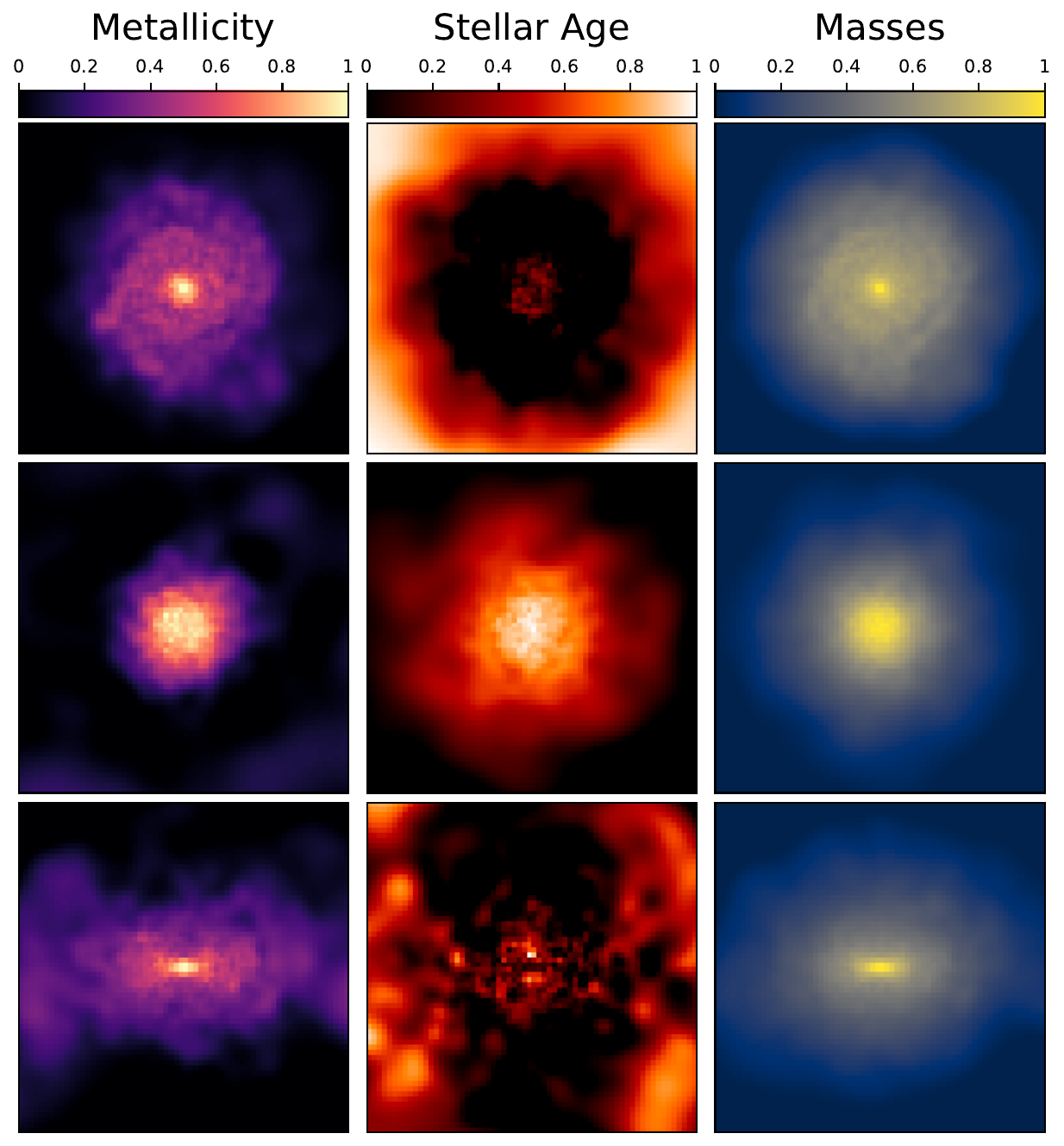}
   }
   \subfigure[3D]{
      \includegraphics[width=0.51\hsize, trim={0 0 2.3cm 0},clip]{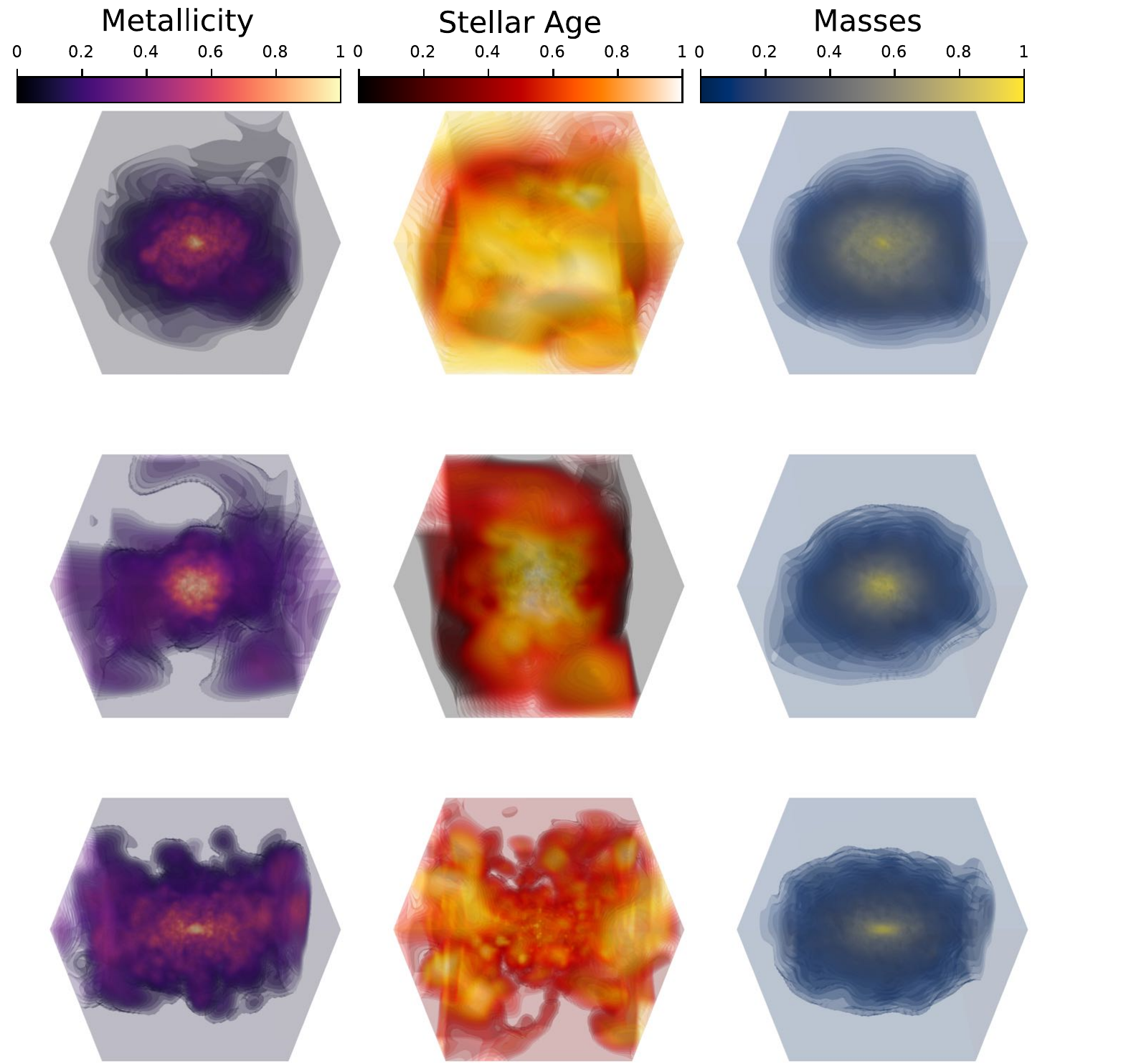}
   }
   \caption{Galaxy images (upper panels) and image cubes (lower panels) in our dataset after the preprocessing steps as explained in Section \ref{sec:dataset_generation}. Each row contains one sample galaxy in three stellar maps - metallicity, age and mass from left to right. The images are normalized in the range $[0,1]$ and bright pixels correspond to high values. Note that in the stellar age maps, dark pixels correspond to young stars. The three dimensional plots have been created using the \emph{plotly} python package.}
   \label{fig: sample_galaxies_after_prepocessing}
\end{figure*}

Hydrodynamical simulations of galaxy formation, such as the large-volume simulations of the IllustrisTNG suite \citep{Pillepich_2017} or the EAGLE suite \citep{Schaye2015} or zoom-in models such as the AURIGA \citep{Grand2017}, FIRE \citep{Hopkins2018}, VINTERGATAN \citep{Agertz2021} or NIHAO \citep{Wang2015,Buck2020} simulations generate rich, multidimensional datasets that include various physical properties of galaxies. These datasets provide an excellent resource for studying galaxy morphology in a more rigorous and quantitative manner. In combination with modern methods of machine learning, these simulations enable us to build an interpretable generative model for galaxy morphology \citep[e.g.][]{Lanusse2021}.    

Fast and accurate models for galaxy morphology are not only used to classify galaxies along the \citet{Hubble_1926} sequence but more recently have gained a lot of attention due to the ESA EUCLID mission. Reliable and accurate models for galaxy images are needed to achieve one of the mission's science goals to measure the weak lensing signal \citep[e.g.][]{Euclid_shape2022}. 

In this work, we set out to study principal component analysis (PCA), a relatively simple, yet powerful, and interpretable model for galaxy morphology. PCA has been widely used in various fields for dimensionality reduction and feature extraction \citep{jolliffe_principal_2016_pca_review}. In \cite{Turk_Pentland_eigenfaces_1991} PCA was used to decompose face images into the subspace spanned by image eigenvectors called "eigenfaces", such that each face can be represented by a weighted sum of the eigenfaces. The term \textit{eigengalaxy} first appeared in \cite{calleja_fuentes_first_eigengalaxies}, where they used PCA for galaxy classification and termed the basis vectors of the lower-dimensional space as eigengalaxies. 

In the context of galaxy morphology, \cite{Uzeirbegovic_2020} recently used PCA to transform a set of SDSS galaxy images into a space where "closeness" corresponds to visual similarity, allowing for a quantitative measure of morphological likeliness.

\begin{figure*}
    \centering
    \includegraphics[width=\hsize]{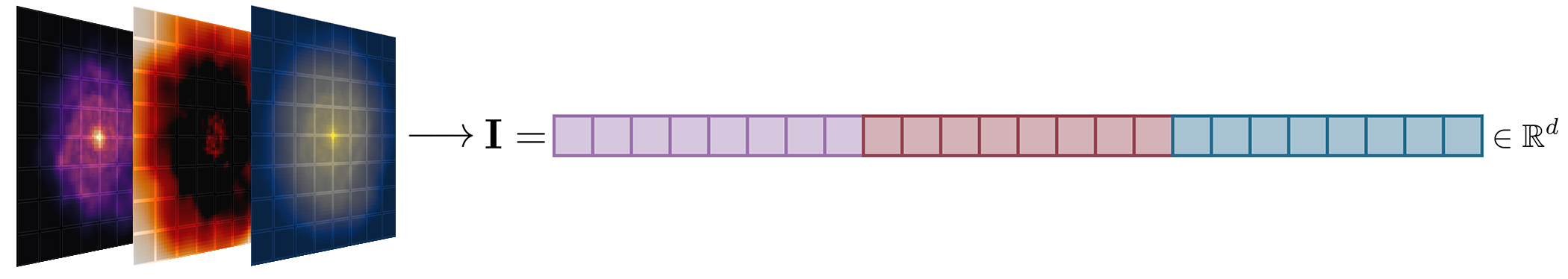}
    \caption{Vectorization in the two dimensional case: This reshaping process converts each galaxy image into a point within a high-dimensional vector space. The first $64^2$ values correspond to the values of the pixels in the metallicity map, followed by the values of the other maps. This allows for a unique vector-based representation of each galaxy as $\mathbf{I} \in \mathbb{R}^d$, where $d$ is the dimension (equation \ref{eq: dimension of images in 2d and 3d}).}
    \label{fig: vectorization of flattened images}
\end{figure*}

Here, we expand upon their findings and explore how we can use PCA to jointly model the distribution of mass, metallicity, and stellar age in two and three dimensions. This approach specifically aims to condense a high-dimensional data distribution from galaxy simulations into a lower-dimensional space that preserves the morphological features of galaxies and allows for interpretable analysis. Our final model is able to jointly model the distribution of stellar mass, metallicity, and stellar age. In combination with a simple stellar population model to convert \U{these parameters} into stellar spectra, it is therefore ideally suited as an interpretable physics model to reconstruct the fundamental joint distribution of \U{these properties} from broad-band galaxy images.

This paper is structured as follows. In Section \ref{sec:methods} we detail our methods. Especially, Section \ref{sec:dataset_generation} discusses the methodology used for generating the dataset for our analysis by extracting galaxy data from simulations. Section \ref{sec: methods_PCA_on_images} outlines the application of PCA to the image data and the resulting lower-dimensional image space (Section \ref{sec: dimensionality reduction method using pca}). Our resulting model is presented in Section \ref{sec: Results} and its accuracy thoroughly tested in Section \ref{sec: model evaluation, reconstruction error}. In Section \ref{sec: application} we highlight potential applications of our PCA galaxy model and conclude in Section \ref{sec: conclusion} with a summary and outlook of our results.

Finally, we publicly release all of our code to reproduce the results of this manuscript via GitHub\footnote{URL: {\url{https://github.com/ufuk-cakir/MEGS}}}and refer to Appendix \ref{sec:appendix_code_and_data} and Fig. \ref{fig: hdf5_file_struc} for an overview of our code and file structure. Our final dataset is publicly available on Zenodo.\footnote{URL: \url{https://zenodo.org/record/8375344}}

%--------------------------------------------------------------------
\section{Methods}
\label{sec:methods}
This section outlines the methodology used to generate the dataset and build the morphology model used in this research.

\subsection{\U{Motivation}}
\label{sec:motivation}
\U{In this paper we set out to explore the possibilities of PCA to condense a high-dimensional data distribution from galaxy simulations into a lower-dimensional space that preserves key morphological features of galaxies and allows to be employed as a flexible, generative model for galaxy morphology informed by high-resolution state-of-the-art simulations. Our final method is able to project the high-dimensional joint distribution of stellar mass, metallicity, and stellar age into a lower dimensional representation whch can easily be used for downstream generative modelling of galaxies. For example, in combination with a simple stellar population models to convert mass, metallicity and stellar age into stellar spectra, it is therefore ideally suited as an interpretable, generative physics model to reconstruct the fundamental joint distribution of these properties from broad-band galaxy images.
Hence, our approach here serves the purpose of tackling the inverse problem of reconstructing physical parameters from observed data.
}

\subsection{From Simulation to Dataset: Galaxy Image Generation}
\label{sec:dataset_generation}
We extracted galaxy data from the IllustrisTNG simulation suite \citep{Nelson_2017,Pillepich_2017,Springel_2017}, specifically using the TNG100-1 simulation, which models galaxy formation in the cosmological context and follows $1820^3$ resolution elements from redshift $127$ to $0$. From this dataset we choose snapshot number $99$ since it corresponds to redshift $z=0$ and extract the galaxies for our later analysis as follows:
\begin{enumerate}
  \item \textbf{Selection} We choose galaxies within a specified mass range of $10^{9.5}M_\odot/h < M_\star < 10^{13}M_\odot/h$, excluding galaxies with \texttt{SubhaloFlag}$=0$, since those are not thought to be of cosmological origin. The final selection yields a total number of \U{$N=\num{11960}$} galaxies in the dataset.
   \\
  \item \textbf{Rotation}: To achieve uniform orientation and meaningful comparisons throughout the data set, we calculated the eigenvectors of the moment of inertia tensor to perform a face-on rotation in the $x-y$ plane and obtain a set of rotated coordinates. Using these coordinates, a temporary image is generated, and a principal component analysis with two components is applied. The first principal component represents a vector that maximizes the variance of the projected particle coordinates\citep[see][]{Uzeirbegovic_2020}. From this component, we determine the tilt angle and further rotate the galaxy to align its semimajor axis with the x-axis. This alignment procedure ensures consistent orientation among elliptical and barred spiral galaxies, thereby enabling reliable comparisons.\\

\item \textbf{Image Rendering}: To accurately capture the visual appearance of the galaxies and reflect the physical, spatial resolution of the simulation, we used the smoothing-length parameter to generate two- and three-dimensional images. Specifically, we render images in the fields of \texttt{particle mass, mass-weighted metallicity}, and \texttt{mass-weighted stellar age}\footnote{In the IllustrisTNG catalogue these correspond to the fields: \texttt{Masses}, \texttt{GFM\_Metallicity} and \texttt{GFM\_StellarFormationTime}. Note that we convert the time the star was born to an age using the age of the universe, so low values correspond to young stars.} This rendering process is performed using the render module from the \textit{SWIFTsimIO} library \citep{Borrow2020}. We choose an image resolution of $64 \times 64$ and $64\times 64\times 64$ pixels for the two- and three-dimensional case, and the image section to be rendered spans five stellar half-mass radii ($5R_{\text{half}}$) in each direction. \U{The resolution was chosen to correspond to the native resolution of large observational integral field spectroscopic datasets such as e.g. MANGA \citep[75\% of MANGA cubes have less than 64x64 spaxels][]{Bundy2015} or SAMI \citep[50x50 spaxels][]{Allen2015} for which we have in mind that our model will be most useful.}\\

\item \textbf{Normalization and Clipping}: Normalization involves rescaling the pixel values to $[0, 1]$, which standardizes the intensity levels across the images, allowing comparisons between different galaxy images. Clipping is employed to remove noise by setting a threshold for pixel values, i.e. we discard values that deviate significantly from the majority of pixels. We find that clipping values below the $25^\text{th}$ percentile works best. This helps to highlight important features while reducing the influence of irrelevant variations \U{which mainly arise from shot noise from the limited particle count in individual pixels originating from the limited resolution of the underlying simulations}.

The normalization and clipping steps are crucial for our subsequent analysis. By eliminating noise and standardizing the pixel values, we facilitate the identification and extraction of meaningful information about galaxy properties, such as morphological characteristics, stellar age distribution, and metallicity patterns. This is also needed for better PCA decomposition by removing the nuisance variance originating from random noise.
\end{enumerate}

We iterate over all selected galaxies and computed both two- and three-dimensional images for the three specified fields. The resulting images are then saved in the HDF5 file format, encompassing the entire dataset. In Fig.~\ref{fig: sample_galaxies_after_prepocessing} we show three representative examples of galaxies (face-on spiral, elliptical, and triaxial) in both \U{2}D (upper panels) and 3D (lower panels). In each subfigure we show all three maps (metallicity, age, and mass from left to right). In addition to 2D images and 3D cubes, we store the SubhaloID \U{(i.e the unique identifier of the galaxy in the simulations)} and stellar mass values $M_\star$ for each galaxy in the \textit{ attributes} subgroup. It is important to note that in the case of IllustrisTNG, we exclude particles with negative \texttt{GFM\_StellarFormationTime}, because such particles correspond to gas cells in a wind phase and are not considered relevant for the analysis. The final data set used in this work is further described in detail in \cite{Cakir2023} and can be found publicly available here: \href{https://github.com/ufuk- cakir/GAMMA}{https://github.com/ufuk- cakir/GAMMA}

\subsection{Principal Component Analysis on Images}
\label{sec: methods_PCA_on_images}

Principal Component Analysis (PCA) is a powerful mathematical technique commonly used in data analysis and dimensionality reduction. In simple terms, PCA finds a new set of basis vectors (\emph{principal components}) that maximizes the variance of the projected data in each direction. The principal components are ordered by their contribution to the total variance of the data, so the first principal component is the direction of maximal variance. Then, the dimensionality of the data space can be reduced by keeping only the top $n$ principal components and projecting all data points onto these principal components to find a lower dimensional representation.

In the context of image analysis, PCA offers a rather simple yet powerful approach to extracting not only meaningful features from images and understanding the underlying structures in an unsupervised manner, but also to performing this in a fast and interpretable way.

%\subsubsection{Building the Data Matrix from Galaxy Images}
%\label{sec: methods_building_data_matrix}

In order to perform PCA, we first need to construct a proper data matrix. Each image in our dataset is represented as a two-[\textbf{three-}] dimensional array of size $64\times64$ [ \U{$64\times64 \times 64$}] with values ranging from $0$ to $1$. For each galaxy, the images have been calculated on three different maps (as explained in Section \ref{sec:dataset_generation}).
These images can be reshaped into a single row vector (see Fig. \ref{fig: vectorization of flattened images}), transforming each galaxy into a point within a high-dimensional vector space, denoted as $\mathbf{I} \in \mathbb{R}^d$, where the dimensionality $d$ is equal to the total number of pixels in the three combined maps. This results in a dimension of
\begin{equation}
\begin{split}
    d_{2D}&=3\cdot64^2 = \num{12288} \\
    d_{3D}&=3\cdot64^3=\mathbf{\num{786432}}
\end{split}
    \label{eq: dimension of images in 2d and 3d}
\end{equation}
in the two and three dimensional case. Note that the order in which the three different fields are flattened is not significant as long as it remains consistent for all the galaxies. This consistency is crucial because we focus on examining the pairwise relationship between each pixel in the high-dimensional space. We finally concatenate all flattened galaxy row vectors to form our data matrix
\begin{equation}
\mathbf{D}=\left[\begin{array}{c}
\mathbf{I}_1 \\
\mathbf{I}_2 \\
\vdots \\
\mathbf{I}_N
\end{array}\right] \in  \mathbb{R}^{N\times d}
\label{eq: datamatrix_definition}
\end{equation}
with $N=\num{11960}$ (number of galaxies in the dataset) and $d = d_{2D},d_{3D}$.

\subsubsection{Dimensionality Reduction}
\label{sec: dimensionality reduction method using pca}
Our goal is to reduce the dimensionality of the original image space while retaining as much information as possible. PCA addresses this by projecting the data onto a new set of orthogonal axes that maximize the variance of the projected data.

%$(11960; 3\cdot64^2)$ [$\mathbf{(11960; 3\cdot64^3)}$]
We perform PCA on our data matrix $\mathbf{D} \in \mathbb{R}^{N\times d}$ using the \emph{scikit-learn } \citep{scikit-learn} python package.
 The PCA method implements the following steps in a computationally efficient way:
\begin{enumerate} 
    \item \textbf{Centering}: Subtract the mean galaxy image to obtain the column-centered data matrix $\mathbf{D}^*_{ij}$:
    \begin{equation}
        \mathbf{D}^*_{ij} = \mathbf{D}_{ij} - \mathbf{\mu}_j
        \label{eq: centering the data}
    \end{equation}
    where $\mathbf{\mu} \in \mathbb{R}^d $ is the column-wise mean of the data matrix $\mathbf{D}$ The mean images used to center the data are shown in \U{ Fig. \ref{fig:mean galaxy images}}.\\
    
    \item \textbf{Calculate SVD}: 
    Singular Value Decomposition (SVD) decomposes the centered data matrix into a product of three matrices
    \begin{equation*}
        \mathbf{D}^* = \mathbf{U} \mathbf{\Sigma} \mathbf{V}^T
    \end{equation*}
where $\mathbf{U} \in \mathbb{R}^{N\times N}$, $\mathbf{V} \in \mathbb{R}^{d\times d}$ are unitary matrices, whose columns represent the eigenvectors of $\mathbf{D}^* \mathbf{D}^{*T}$ and $\mathbf{D}^{*T} \mathbf{D}^*$ respectively. The rectangular matrix $\mathbf{\Sigma}\in \mathbb{R}^{N\times d}$ contains the singular values of $\mathbf{D}^*$, which are the square roots of the eigenvalues of $\mathbf{D}^{*T} \mathbf{D}^*$\\
\end{enumerate}
The eigenvectors correspond to the principal components of our data and are ordered by their eigenvalues, which represent the amount of variance explained by each component. The eigenvectors $\mathbf{\lambda} \in \mathbb{R}^d$ represent new orthogonal basis vectors that span our lower-dimensional image space. These eigenvectors are called \emph{eigengalaxies}, since they can be interpreted as images.

\begin{figure}
    \includegraphics[width=\hsize]{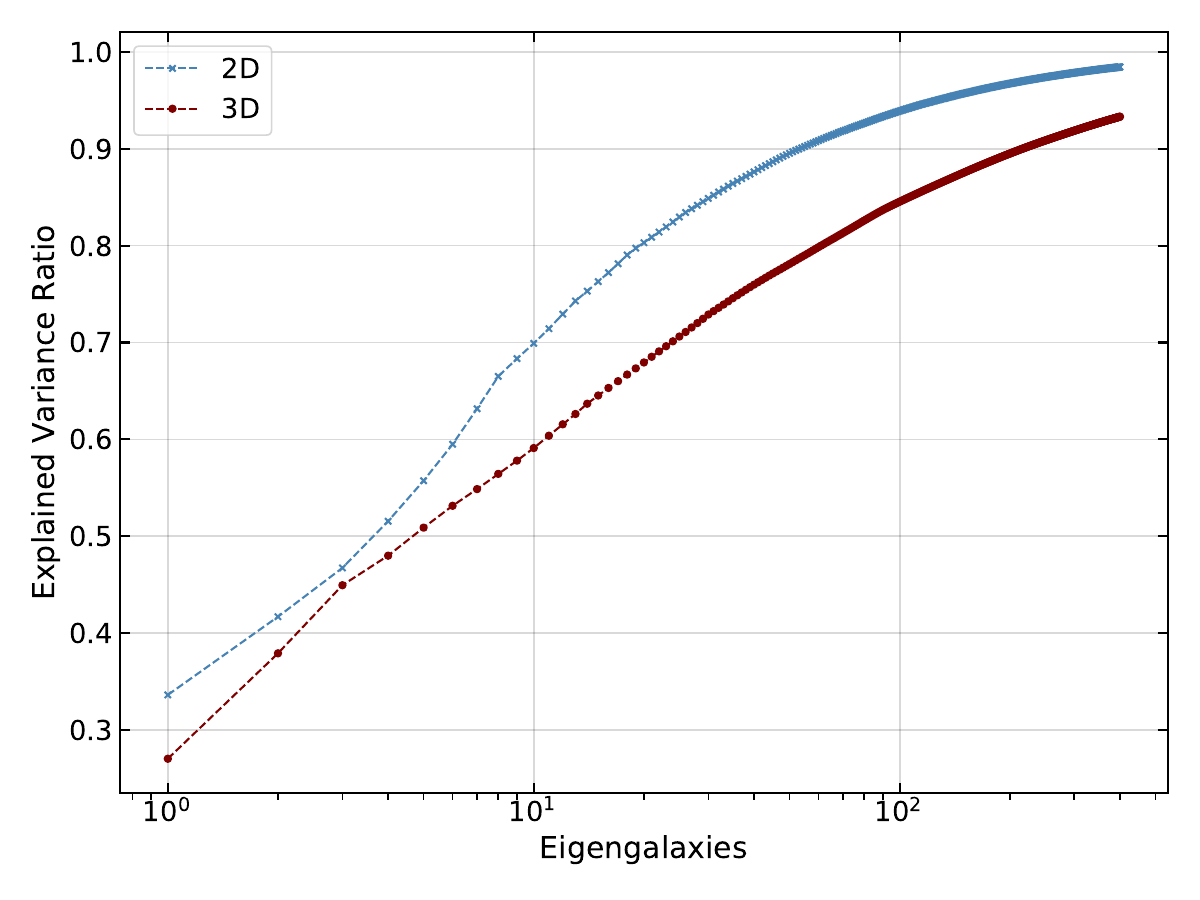}
    \caption{Cumulative sum of explained variance ratio for up to 400 eigengalaxies (left): We observe that one needs much more eigengalaxies in three dimensions to achieve the same explained variance as in two dimensions. This Figure shows that in order to achieve more than 90\% EVR, you need around 60 (215) eigengalaxies in two (three) dimensions.}   
    \label{fig:cumsum_explained_variance_ratio}
\end{figure}

\begin{figure}
    \centering
    \includegraphics[width=\hsize]{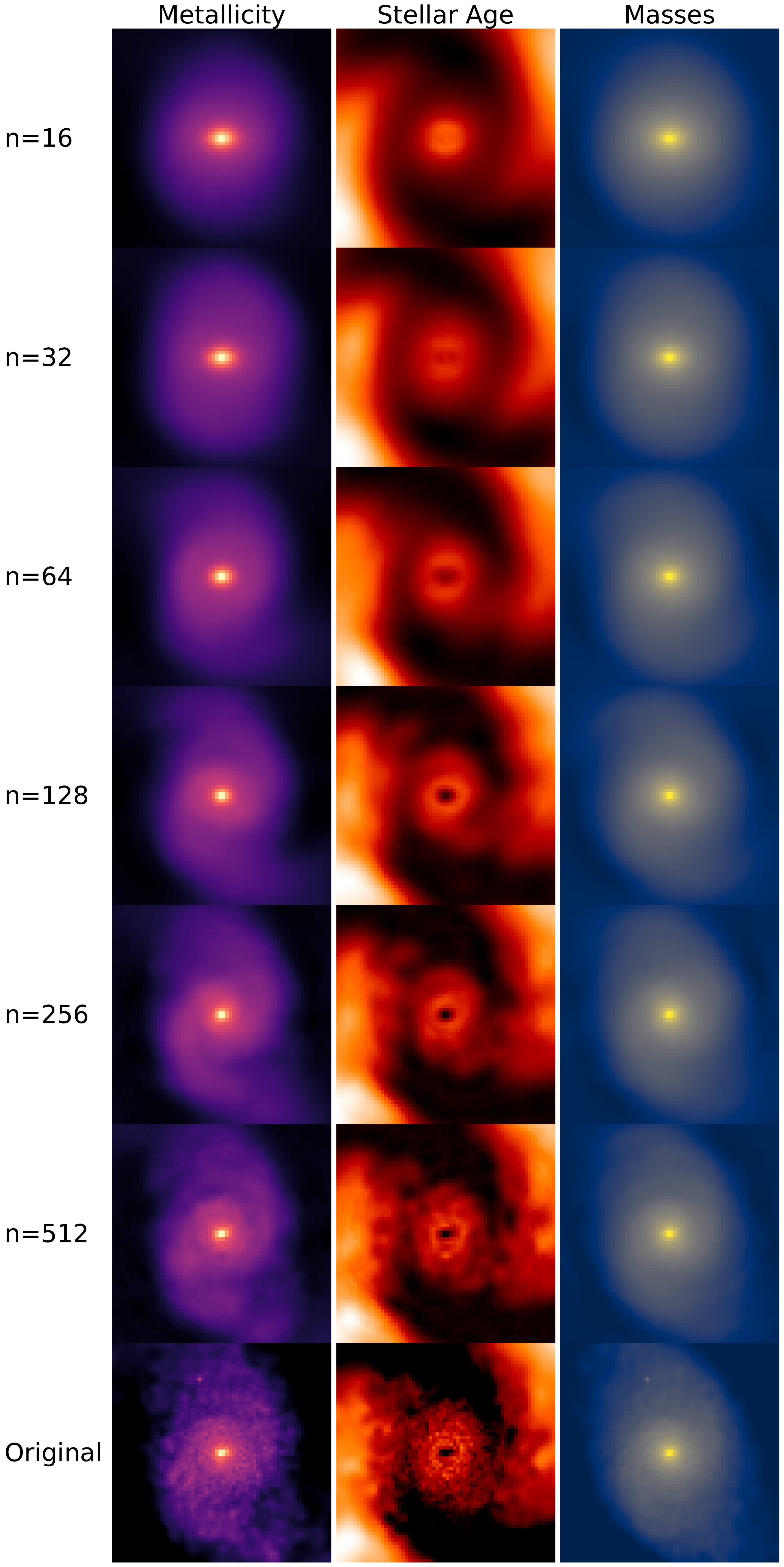}
    \caption{Each column shows the lower dimensional representation using $n$ eigengalaxies. To highlight how close we get, the original images are shown at the bottom. One can see that higher order eigengalaxies account for more detailed small scale structures, however only 16 eigengalaxies are enough to fit the overall morphology of the galaxy.}
    \label{fig:reconstruciton with with different n eigengalaxies}
\end{figure}

 \begin{figure*}
   \centering
   \subfigure[Metallicity]{
      \includegraphics[width=.95\hsize]{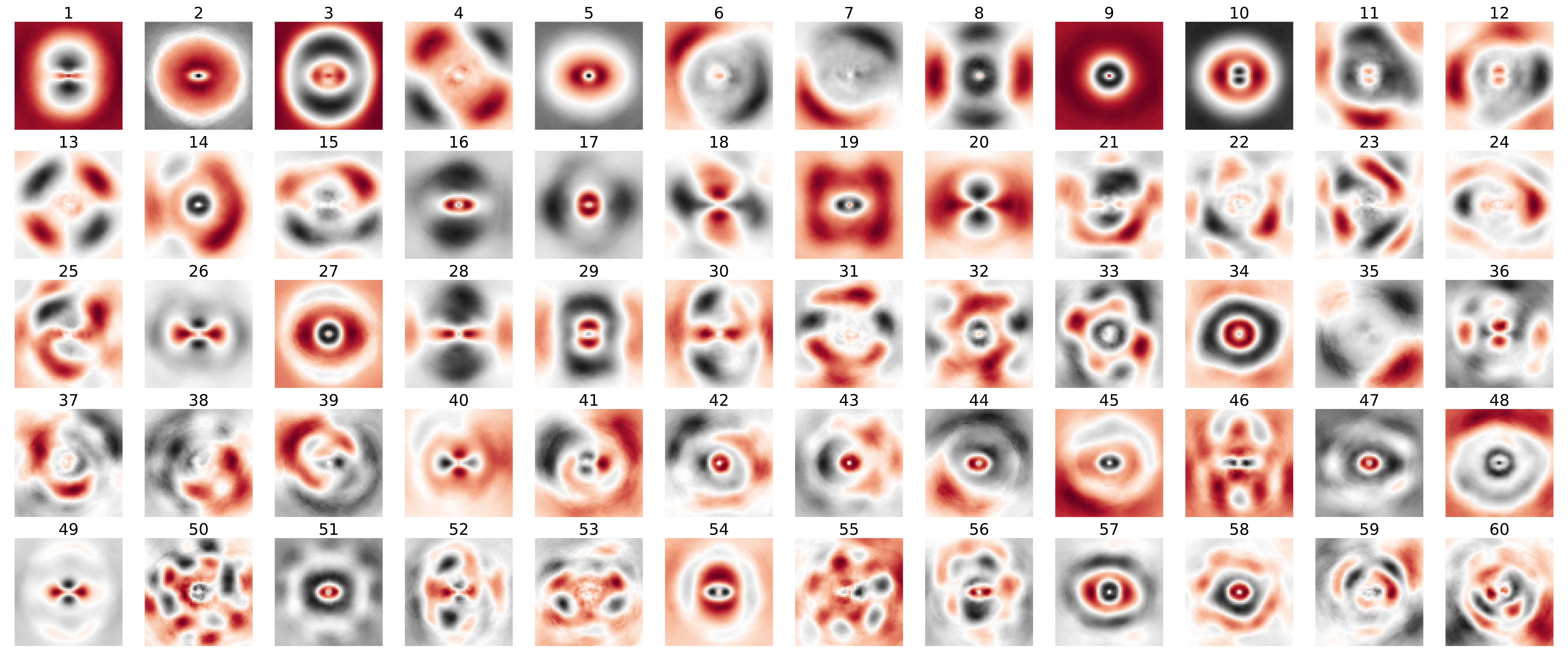}
   }
   \subfigure[Stellar Age]{
      \includegraphics[width=.95\hsize]{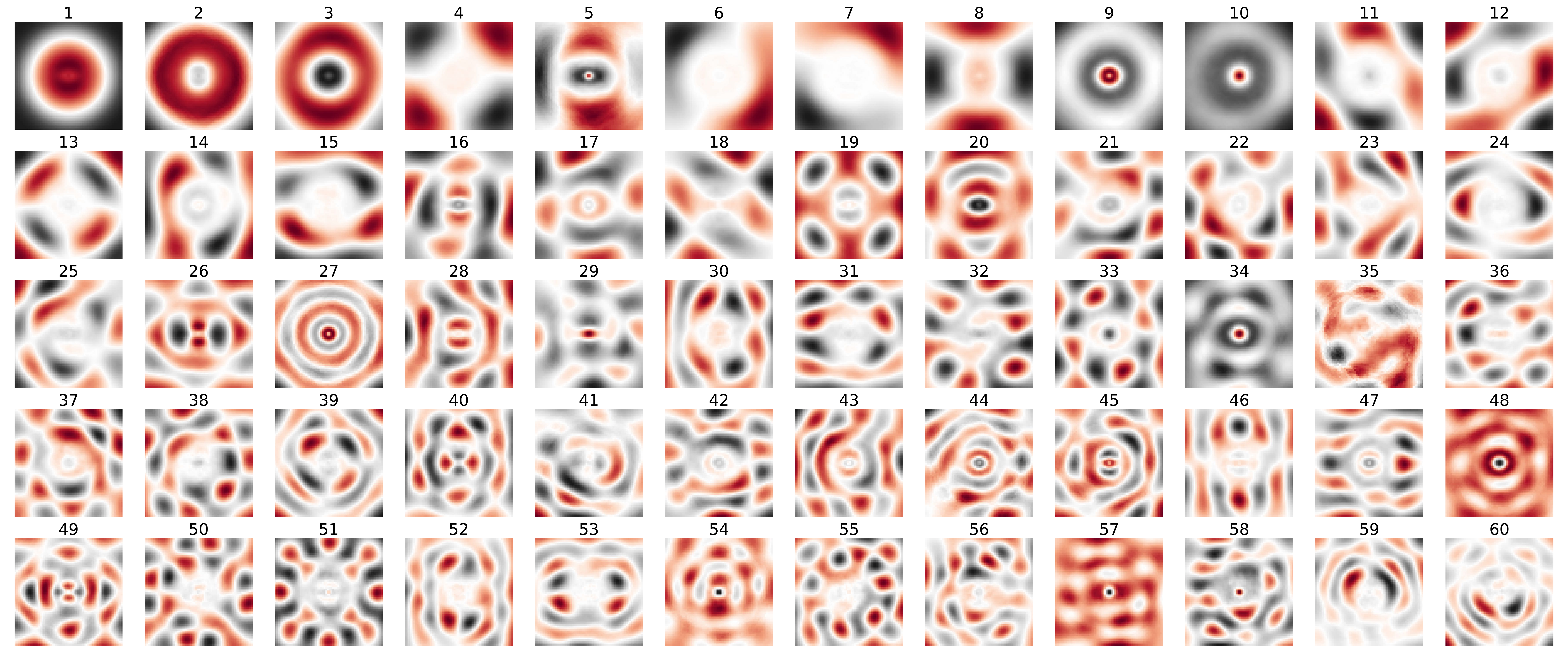}
   }
   \subfigure[Masses]{
      \includegraphics[width=.95\hsize]{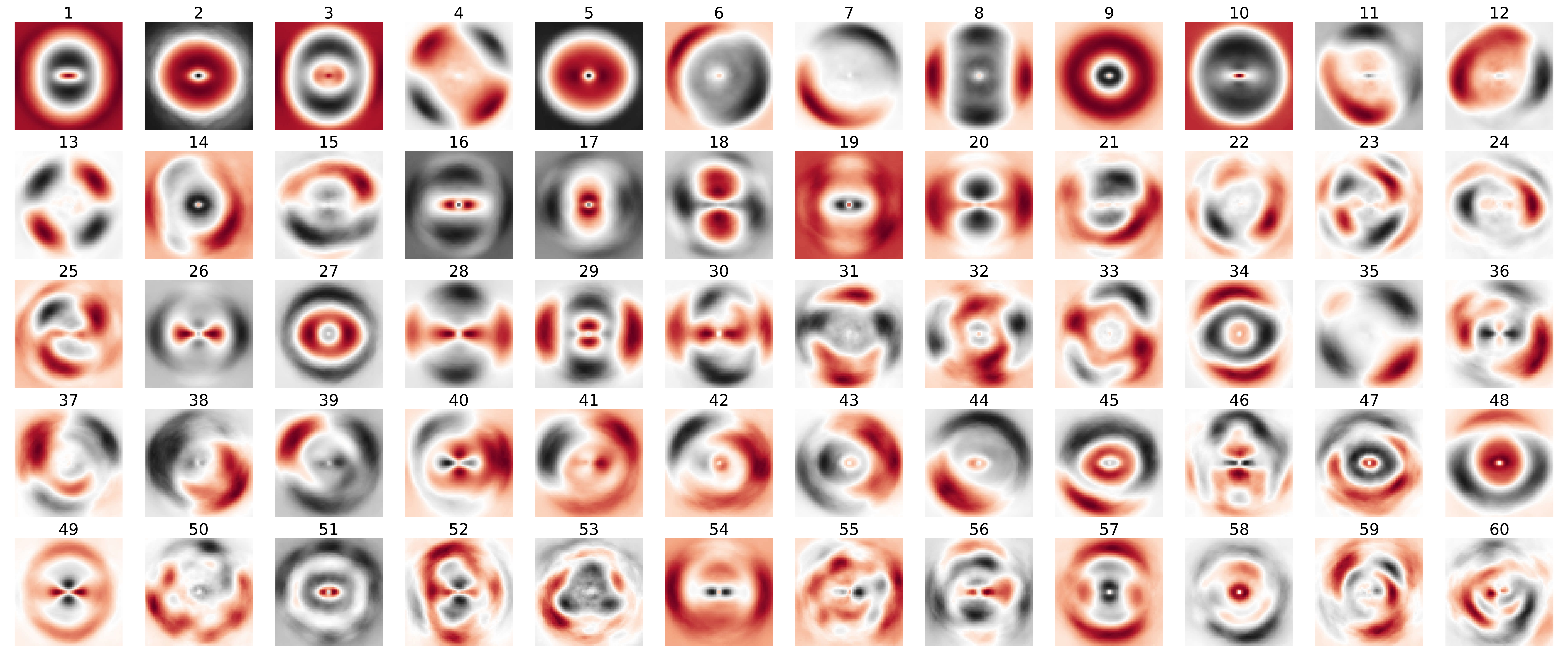}
   }
   \caption{First 60 Eigengalaxies from PCA in 2D: Red corresponds to high positive values, and gray corresponds to low negative values. We show the three different eigengalaxies for metallicity, stellar age, and mass from top to bottom.}
   \label{fig: eigengalaxies 2D}
\end{figure*}   

We project our high-dimensional data matrix $\mathbf{D}^*$ onto this subspace spanned by the top $n$ eigengalaxies $\mathbf{\lambda_n} \in \mathbb{R}^{n\times d}$ to obtain a lower-dimensional representation.  
\begin{equation}
    \mathbf{S} = \mathbf{D}^*\lambda_n^T 
    \label{eq: pca_scores_projection}
\end{equation}
where $\mathbf{S} \in \mathbb{R}^{N\times n}$ contains the coordinates of $N$ galaxy samples in the $n$-dimensional eigengalaxy subspace, which we also refer to as the \emph{PCA scores}. This representation maintains most of the variance in the data while also reducing the dimension from $d$ to $n$.

In order to transform the scores back to the original image space, we simply calculate the data matrix $\hat{\mathbf{D}}$ containing the PCA reconstructed galaxy images as the weighted sum of $n$-eigengalaxies:
\begin{equation}
    \hat{\mathbf{D}} = \mathbf{S} \mathbf{\lambda_n} +\mathbf{\mu} 
    \label{eq: galaxy reconstruction formula}
\end{equation}

\section{Results}
\label{sec: Results}

The PCA-based morphology model provides a quantitative analysis of galaxy morphology by examining the positions of galaxies in the lower dimensional space to study morphological similarities and to model the joint distribution of mass, metallicity and stellar age. In this section we present our final PCA decomposition of galaxies in 2D and 3D and evaluate its performance via several quantitative metrics.

\subsection{Dimensionality Reduction and Explained Variance}
\label{sec: dimensionality reduction}

An important quantity to evaluate how good the PCA model preserves information is the explained variance, which refers to the amount of variance of the data that is captured by each principal component, i.e., eigengalaxy. The Explained Variance Ratio (EVR) for each eigengalaxy is the proportion of the total variance of the data set that is explained by that component. Since the eigengalaxies are orthogonal, each additional eigengalaxiy accounts for variance not explained by the eigengalaxies before.
%In section \ref{sec: dimensionality reduction} we explore the dependence of the reconstruction accuracy and explained variance on the number $n$ of eigengalaxies used.

We are free to choose the total number $n$ of eigengalaxies we want to keep for the lower-dimensional image space representation (Section \ref{sec: dimensionality reduction method using pca}). We calculate the cumulative sum of the EVR of each eigengalaxy in Fig. \ref{fig:cumsum_explained_variance_ratio} for up to 400 components and find that, e.g. 60 (215) eigengalaxies account for about 90\%  of the total variance in 2D (3D), which means that most of the information from the dataset can be retained even after significant dimensionality reduction (e.g. a reduction by a factor of $205$ ($3641$)). 

PCA gets better with increasing number of components, however one needs to decide for the trade-off between reconstruction accuracy and dimensionality reduction.
This is showcased in Fig. \ref{fig:reconstruciton with with different n eigengalaxies} which contains the lower-dimensional representation of a sample galaxy (eq. \ref{eq: galaxy reconstruction formula}) using different numbers of eigengalaxies. From top to bottom, we show the reconstructed galaxy image using 16 to 512 eigengalaxies while the bottom row shows the original image for comparison. Already with 16 eigengalaxies the main features of the galaxy are well reconstructed, especially the spiral-like structure of stellar age. In general this figure clearly shows that with increasing number of eigengalaxies used for the reconstruction the image accuracy or the finer image details become more and more well approximated. We note that between 256 and 512 eigengalaxies we mostly observe that the noise level in the image is better approximated.

The choice of how many eigengalaxies to keep is heavily dependent on the use case, however for further analysis in this paper we use 60 (215) eigengalaxies in two (three) dimensions, since those account for 90\% of explained variance as we have shown in Fig. \ref{fig:cumsum_explained_variance_ratio}. The resulting 60 eigengalaxies in the 2D case are shown in Fig.~\ref{fig: eigengalaxies 2D} and the corresponding first 60 eigengalaxies of the 3D case are shown in the appendix in Fig.~\ref{fig: eigengalaxies 3D}. We observe that the structures get much more complex as we go up to higher orders up to the scale where we observe pure rotational noise, especially visible in the case of stellar age eigenmaps. This is explained by the fact that the eigengalaxies are ordered by explained variance, and the lower orders account for more global structures. \U{By rotational noise, we mainly refer to the fact, that galaxies show spiral arm structure and other patterns with rotational symmetry. Since PCA is a linear dimensionality reduction, it is not able to faithfully capture these features. Hence higher order eigengalaxies are needed to account for these details. This also points towards a limitation of our model and warrants to explore geometric and equivariant models in future work.} 

Interestingly, the physical correlation between mass, metallicity, and age is incorporated by the fact that the stellar age eigengalaxies are almost inverted to those of metallicity and mass, since young stars have on average higher metallicity and are located in overdense regions, i.e. the spiral arms. Note that you even observe eigengalaxies with explicit spiral arms (e.g eigengalaxy 25 in Fig. \ref{fig: eigengalaxies 2D}). Thus we conclude that decomposing the physical appearance of galaxies into eigengalaxies indeed results in an interpretable lower-dimensional representation.  

%To further assess the robustness of the model, we fit PCA with 60 components 400 times on a subsample composed of 75\% randomly selected galaxies from the dataset, to check if the results are significantly influenced by the specific sample of galaxies chosen. For each trial, we record the explained variance and evaluate its distribution (cf.  Fig. \ref{fig:400_evr_calc_60_components} in the Appendix). The result of our bootstrapping approach shows a really good robustness and has a mean of $0.9087$ and a standard deviation of only $0.0003$. From this we conclude that a) our dataset size of $N=\num{11960}$ galaxies is large enough to cover a decent amount of different galaxy morphologies and b) that using a number of 60 eigengalaxies is enough to robustly explain $90\%$ of the variance between different galaxy morphologies.

\begin{figure*}
    %\centering
    \subfigure[2D]{
    \includegraphics[width=.5\hsize]{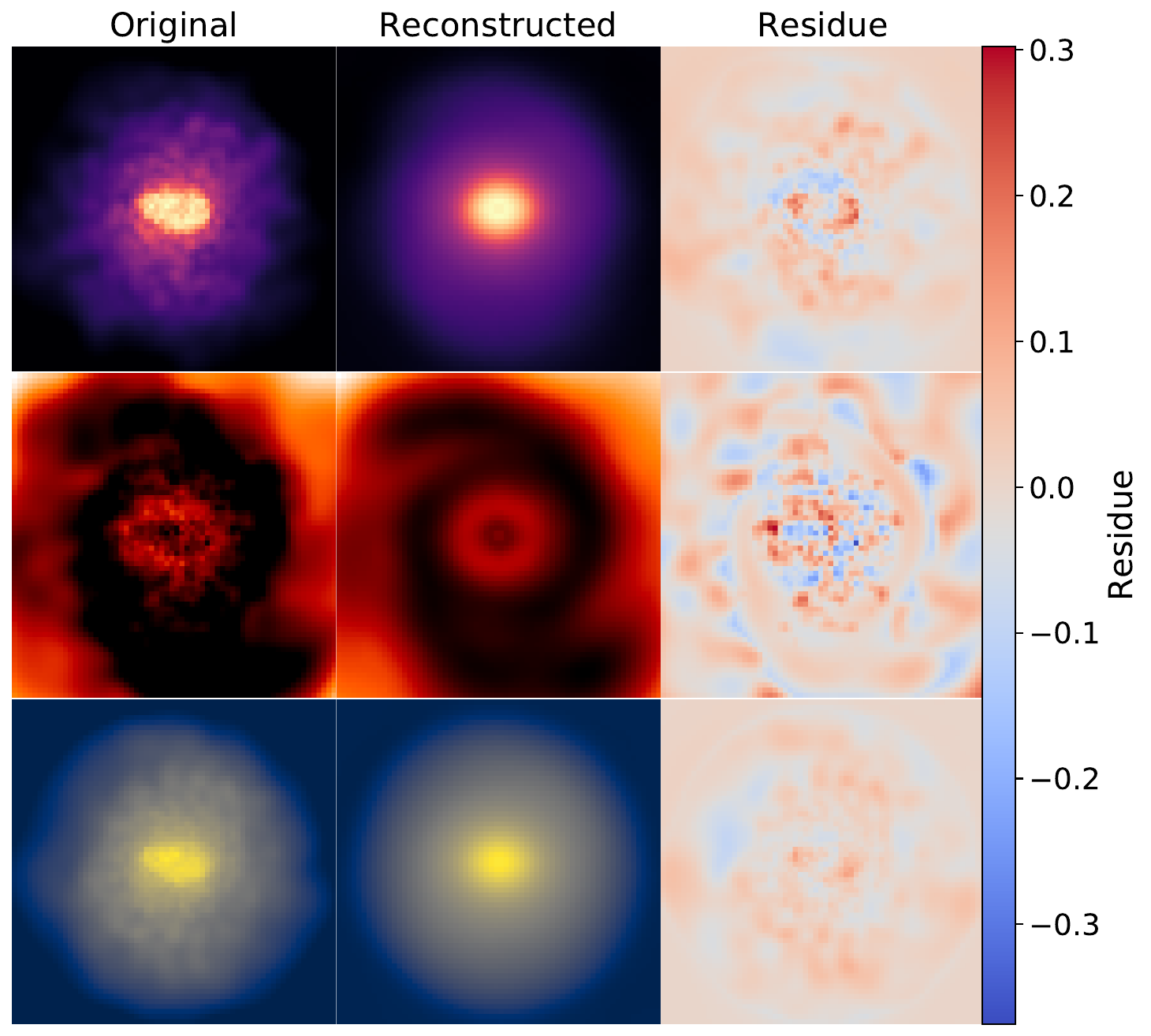}
    }
    \subfigure[3D]{
    \includegraphics[width=.5\hsize,trim={0.6cm .1cm 0 0},clip]{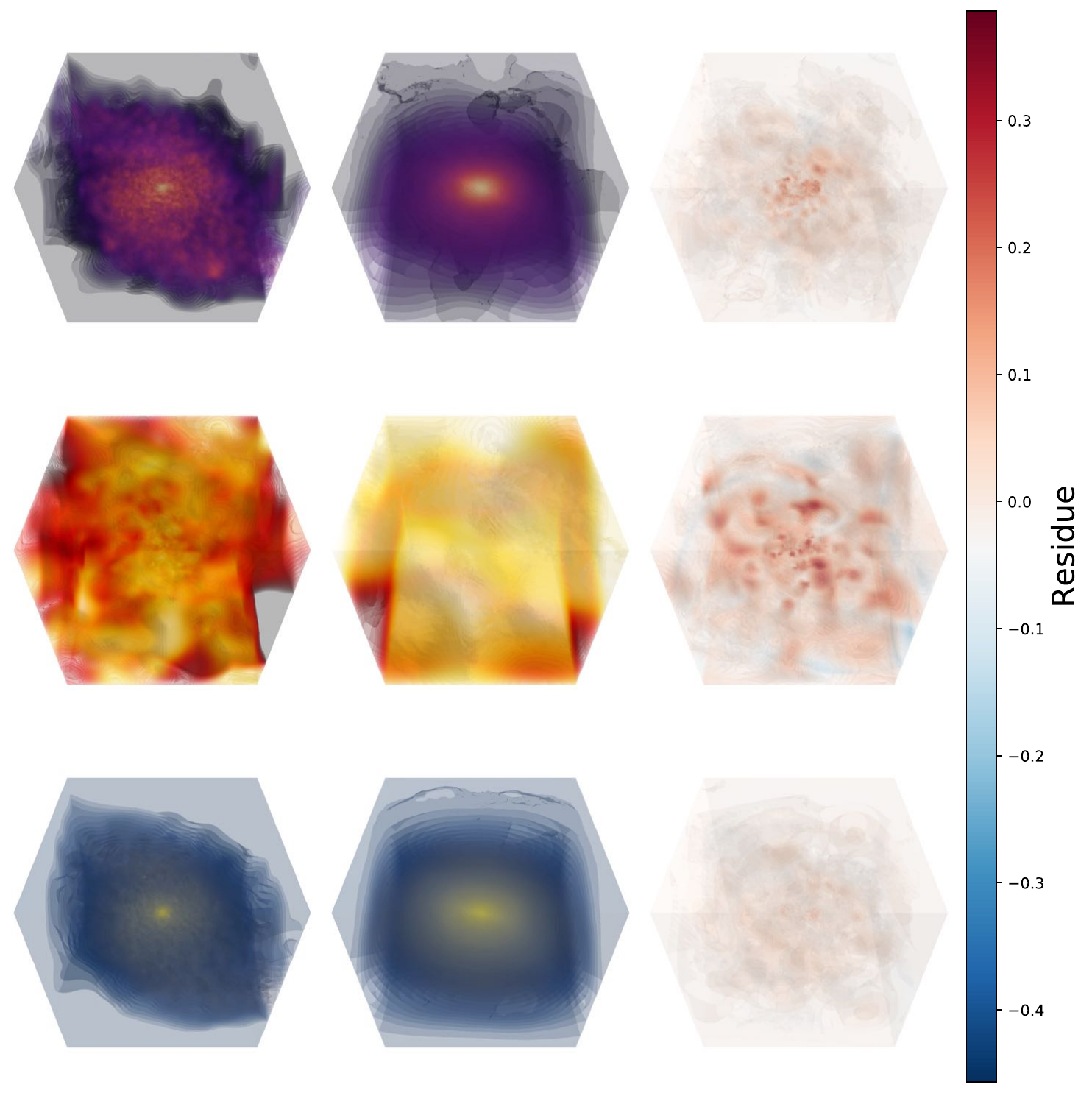}
    }
    \caption{PCA reconstruction in two and three dimensions: The left most column in each panel shows a sample galaxy in the three maps (metallicity, stellar age, and mass from top to bottom) followed by the low-dimensional PCA representation with 60 (215) eigengalaxies in two (three) dimensions. The rightmost column shows the residual image ($\mathbf{I_\textbf{original}} -\mathbf{I_\textbf{rec}}$). You can clearly see that the complex spiral structure in the Stellar Age map is approximated very well in the two-dimensional case while at the same time at $n=60$ we smooth the noise in the image. Compare also with Fig. \ref{fig:reconstruciton with with different n eigengalaxies} for what happens in the larger n.}
    \label{fig:one reconstruction sample galaxy 60 comp 2d}
\end{figure*}

\begin{figure}
    %\centering
    %\subfigure{
    \includegraphics[width=\hsize,trim={0.1cm .1cm 0 0},clip]{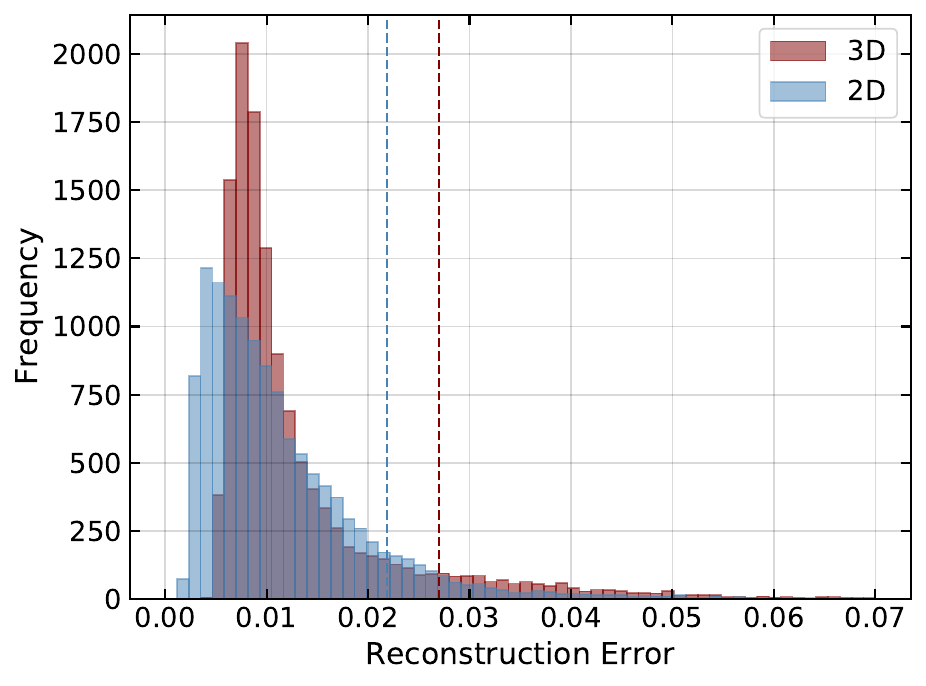}
    %}
    %\subfigure{
    \includegraphics[width=\hsize,trim={0.1 .1cm 0 0.cm},clip]{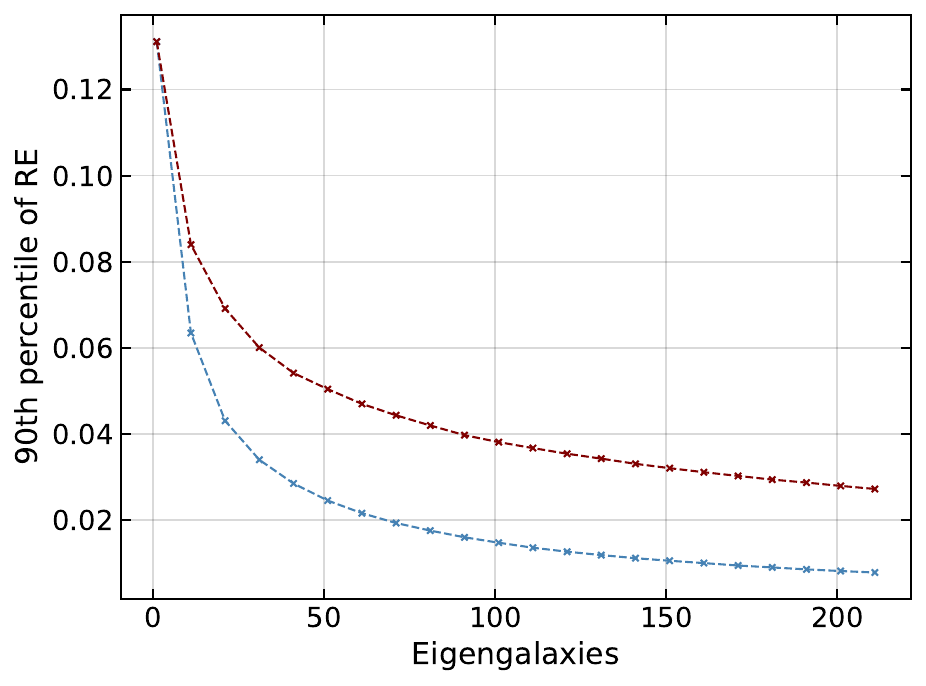}
    %}
    %\caption{\todo{maybe plot the 3d rec err also in here?}Reconstruction Error between original two dimensional images and their PCA projection onto the 60 dimensional space. We observe that 90\% of all images have a reconstruction error less than $0.02$ \T{put the line for 90 percentile as was in the thesis.}\todo{change x range?, exclude extreme values?}\T{yes, I would clip the x-range at 0.06 maybe 0.07. Did you play with log scale for y-axis?}}
    %\label{fig:rec_err_60_comp}    
    \caption{Reconstruction Error between original images and their PCA projection. \emph{Top panel:} Reconstruction error for fixed dimensionality reduction onto the 60 (215) dimensional space in two (\textbf{three}) dimensions. The dashed line represents the 90\% quantile. We observe that 90\% of all images have a reconstruction error less than $0.022$ ($\mathbf{0.027}$). \emph{Bottom panel:} 90th percentile of the reconstruction error as a function of the number of eigengalaxies used for reconstruction. The reconstruction error is a strong function of eigengalaxies, and already 15 (60) eigengalaxies lead to a reconstruction error better than 5\% in 2D (3D).}
    \label{fig:rec_err_60_comp} %\label{fig:rec_err_percentile_vs_eigengalaixes}
\end{figure}

\subsection{Reconstruction Accuracy}
\label{sec: reconstruction accuracy}

With our final set of 60 eigengalaxies for the 2D case, we calculate the lower dimensional representation of all galaxies in our dataset. In the following we denote by $\mathbf{I}_\textbf{original}$ the vector containing the pixel values of the original images and by $\mathbf{I_\textbf{rec}}$ the vector containing the pixel values of the reconstructed image. In our case $\mathbf{I} \in \mathbb{R}^d$ is a $d$ dimensional vector of the image space, where $d$ is given by eq.~\ref{eq: dimension of images in 2d and 3d} for the 2D (3D) case. With this, we define the reconstruction error (RE) as the \U{difference} in pixel values to measure the discrepancy between the PCA representation vector $\hat{\mathbf{I}}$ and the original vector $\mathbf{I}$
\begin{align}
    \text{RE} &= \frac{\sum_{k=1}^{d} (I_k-\hat{I}_k)^2}{\sum_{k=1}^{d} I_k}
    \label{eq:Reconstructin Error}
\end{align}
where $\hat{I}_k$ and $I_k$ represent the $k$-th elements (pixel) of the reconstructed and original vectors (images), respectively.
We calculate the reconstruction error as the \U{difference} in the pixel values between the original and PCA representations separately in the three different maps. An example reconstruction for both the 2D (left panel) and 3D (right panel) case together with the residual image defined as $\mathbf{I_\textbf{original}} -\mathbf{I_\textbf{rec}}$ is shown in Fig. \ref{fig:one reconstruction sample galaxy 60 comp 2d}. Qualitatively, the reconstruction in all three maps is quite well approximating the original image. We find that especially the stellar age map is approximated very well using only 60 components. 

More quantitatively, we calculate the reconstruction error as defined via eq.~\ref{eq:Reconstructin Error} for all galaxies in our sample and show the distribution of reconstruction errors in the top panel of Fig.~\ref{fig:rec_err_60_comp} for both the 2D case (blue histograms) and the 3D case (red histograms). We find that 90\% of all images have a reconstruction error less than $0.022$ in 2D (60 eigengalaxies) and less than $0.027$ in 3D (215 eigengalaxies), highlighting that our PCA model results in accurately reconstructed images. 

Similarly, the bottom panel of Fig.~\ref{fig:rec_err_60_comp} shows how the reconstruction error scales with the number of eigengalaxies used. For the 2D case we find that even with as few as $\sim15$ eigengalaxies 90\% of all images have a reconstruction error below $0.05$. In the 3D case however, we need $\sim 60$ eigengalaxies to achieve the same reconstruction error below $0.05$ which is not surprising as the dimensionality of the problem is much larger. This nicely mirrors the results from the explained variance and shows that indeed only a small number of eigengalaxies are needed to accurately describe galaxy morphology.

 \begin{figure*}
   \centering
   \subfigure{
      \includegraphics[width=\hsize]{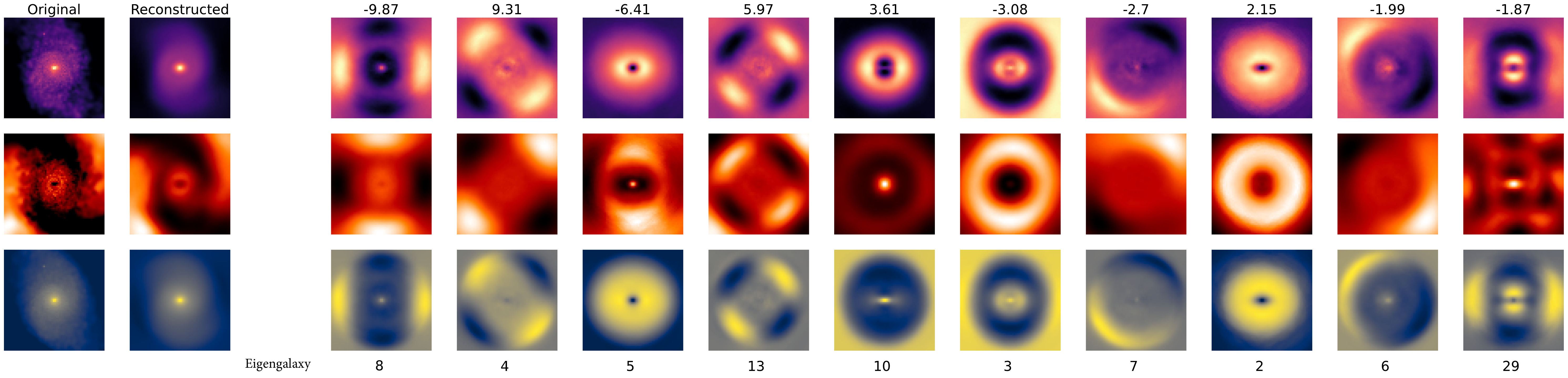}
   }
   \subfigure{
      \includegraphics[width=\hsize]{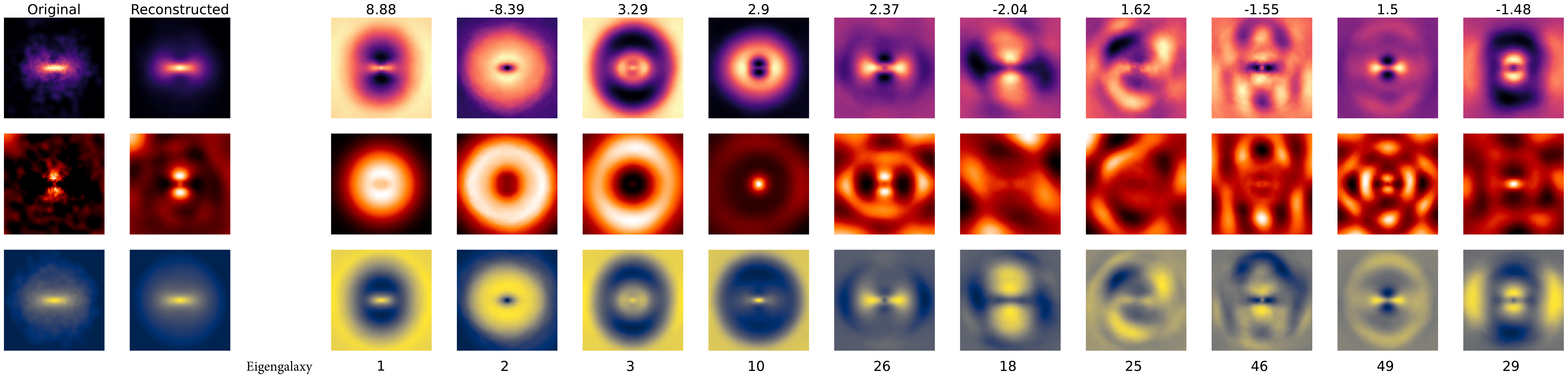}
   }
   \caption{Decomposition of two PCA reconstructed samples: The PCA reconstruction (left) is calculated as the weighted some over 60 eigengalaxies, however here we display the top 10 eigengalaxies (right), chosen based on their absolute weight values (top). \U{On the bottom we show the corresponding index of the eigengalaxy.}. This showcases the different combination of eigengalaxies used to reconstruct galaxies with different morphological features, e.g., spiral or bar shaped}
   \label{fig: decomposition_10}
\end{figure*}

\subsection{Interpretability of the Model}
\label{sec: model evaluation, reconstruction error}

One key advantage of PCA over more advanced machine learning methods such as neural networks is its relatively straightforward interpretability. Since PCA is a linear decomposition of the original image space, each individual component can be interpreted as an image of a very particular morphology. To visualize how eigengalaxies account for different morphology, we plot the 10 eigengalaxies with the highest absolute score values used for reconstruction in Fig. \ref{fig: decomposition_10}. We show here two example cases of the face-on spiral already shown in Fig.~\ref{fig:reconstruciton with with different n eigengalaxies} (upper panels in Fig.~\ref{fig: decomposition_10}) and the triaxial galaxy from the bottom panels in the 2D example from Fig.~\ref{fig: sample_galaxies_after_prepocessing} (lower panels in Fig.~\ref{fig: decomposition_10}).  

The two columns on the left of Fig.~\ref{fig: decomposition_10} show the original and reconstructed images, while the next ten panels show the contributing eigengalaxies in order of decreasing absolute score. The first thing to note is that vastly different eigengalaxies contribute to the reconstruction between a spiral and a triaxial galaxy (only 3 out of 10 eigengalaxies are common, though with very different weights, c.f. 6th vs. 3rd eigengalaxy in the spiral/triaxial case, which have opposite signs), as we would expect from their fundamentally different morphology. For the spiral galaxy, we find eigengalaxies with a lot of power on the diagonal edges of the image, which is needed to reconstruct the spiral arm feature. For the triaxial galaxy, on the other hand, we find that the eigengalaxies with nearly point symmetry or bar-like features contribute most to the reconstruction.

Our analysis of Fig.~\ref{fig: decomposition_10}, although quite qualitative in its nature, clearly shows the potential to describe galaxy morphology through a PCA decomposition. Fitting galaxies with our PCA model will allow one to easily study general morphological trends among different galaxies via the PCA scores of the contributing eigengalaxies (which are all the same for each galaxy by construction). For example, it is easy to select and quantify how bar-like, spiral-like, or centrally concentrated the mass, age, or metallicity distribution of a galaxy is. This further leads to one main application of describing galaxy morphology via PCA, its ability to do galaxy classification and perform a similarity search, which we describe in more depth in the next subsection.

\begin{figure*}
   \centering
   \subfigure[Spiral]{
      
      \includegraphics[width=.4875\hsize]{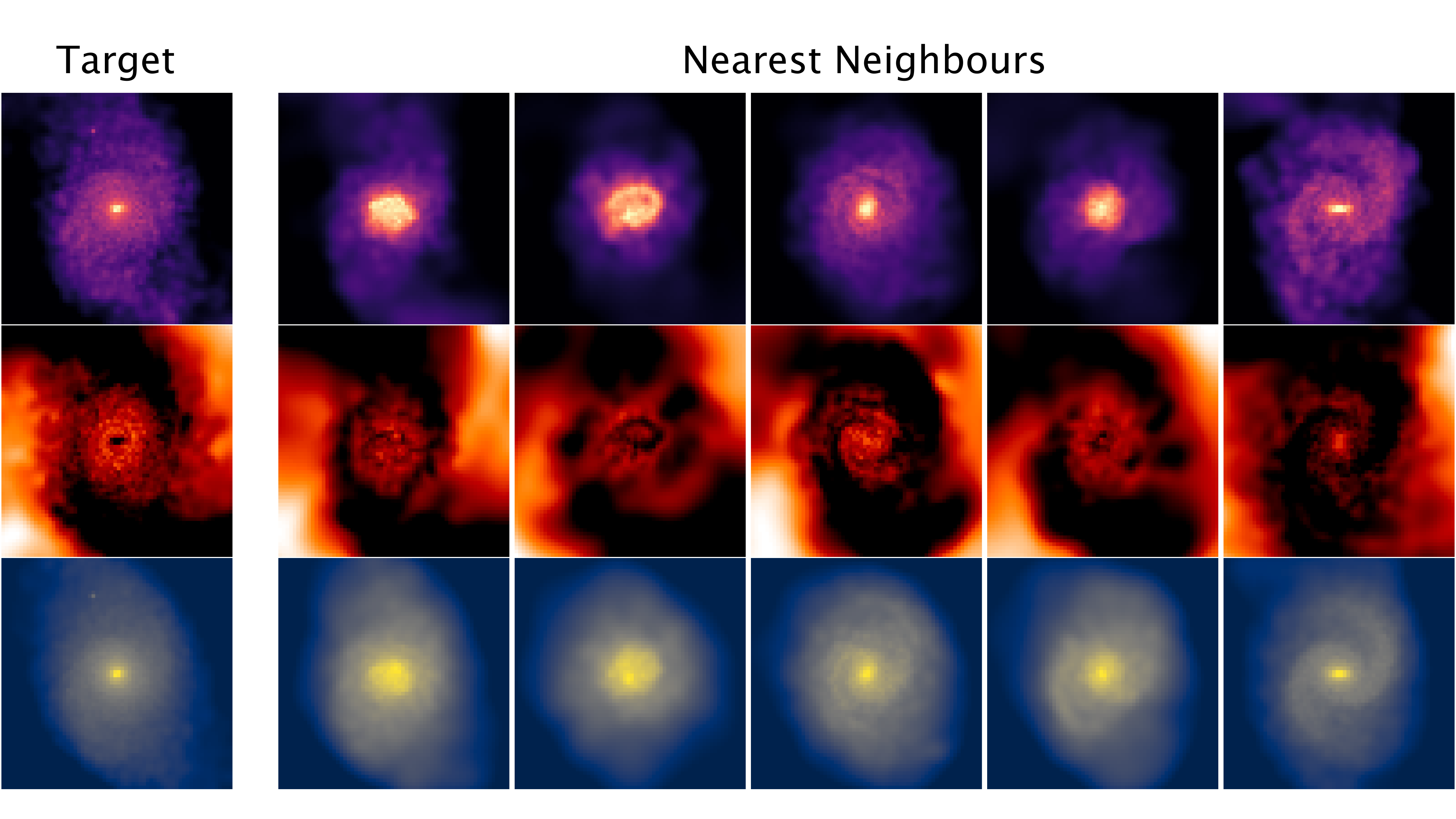}

   }
   \subfigure[Bar]{
      \includegraphics[width=.4875\hsize]{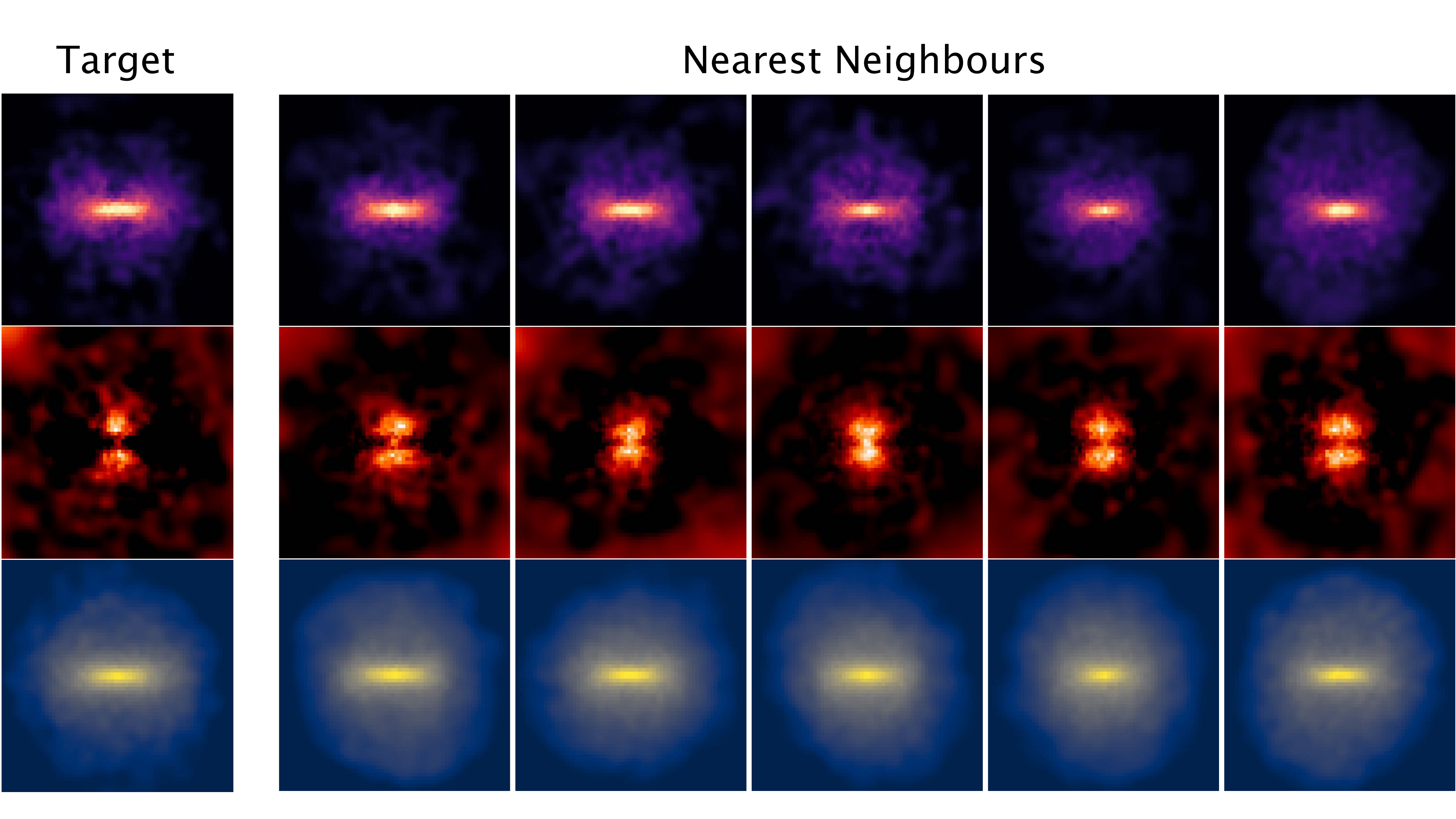}
   }
   \caption{Nearest Neighbours in eigenspace: given a sample galaxy (left) we find the five nearest neighbors in euclidean distance in the lower dimensional space (right) for each a spiral (a) and bar shaped (b) representative galaxy sample.}
   \label{fig: nearest_neighbours}
\end{figure*}

%--------------------------------------------------------------------

\subsection{Application}
\label{sec: application}

In this subsection, we highlight two potential applications of our model: i) a similarity search among galaxies to find morphologically similar galaxies ii) outlier detection via a UMAP projection of the PCA components.
%the reconstruction of fundamental galaxy parameters such as its mass, metallicity, and age distribution from broad-band photometric images. 

\subsubsection{Morphological Similarity Search}

In order to test whether the lower-dimensional image space encodes morphological information, we can select a sample galaxy and search for its nearest neighbors in the PCA eigenspace using the Euclidean distance:
 \begin{equation}
d(\Vec{\hat s}, \Vec{s_i})=\|\Vec{\hat s}-\Vec{s_i} \|_2
\label{eq: euclidean_distance}
\end{equation}
where $\Vec{\hat s}, \Vec{s_i} \in \mathbf{S}$ (eq. \ref{eq: pca_scores_projection}) are the scores of the sample and the remaining galaxies. In Fig.~\ref{fig: nearest_neighbours} we illustrate the five nearest neighbors for the two different sample galaxies already discussed in the previous Fig.~\ref{fig: decomposition_10}.
Indeed, this figure shows clearly that there exists a relationship between Euclidean proximity in the PCA eigenspace and the shared morphological characteristics of galaxies. The five nearest neighbors for the spiral galaxy (left panels of Fig.~\ref{fig: nearest_neighbours}) all \U{generally show} a pronounced two-arm spiral galaxy where the strongest morphological similarity can be seen in the age maps (middle row). The right-hand side panels of Fig.~\ref{fig: nearest_neighbours} show that indeed all neighbors of the triaxial galaxy in the PCA eigenspace are also triaxial and share the same strong bar-like feature in mass and metallicity maps.

Similarly, the PCA scores can further be used to perform clustering analysis (e.g. using gaussian mixture models or k-means) to define galaxy types in an unsupervised fashion or to perform an outlier detection to potentially find some interesting galaxies with morphological particularities. 

\subsubsection{Outlier Detection}
\label{sec: outlier detection}
%We calculate UMAP (Uniform Manifold Approximation and Projection ) \cite{mcinnes2021umap}, a dimensionality reduction and clustering algorithm, on the lower-dimensional PCA scores to visualize the space in two dimensions.

\U{To visualize the space in two dimensions, we apply UMAP (Uniform Manifold Approximation and Projection), a dimensionality reduction and clustering algorithm developed by \cite{mcinnes2021umap}, on the lower-dimensional PCA scores.}

\begin{figure*}
    \centering
    \subfigure[]
    {
    \includegraphics[width = .48\textwidth]{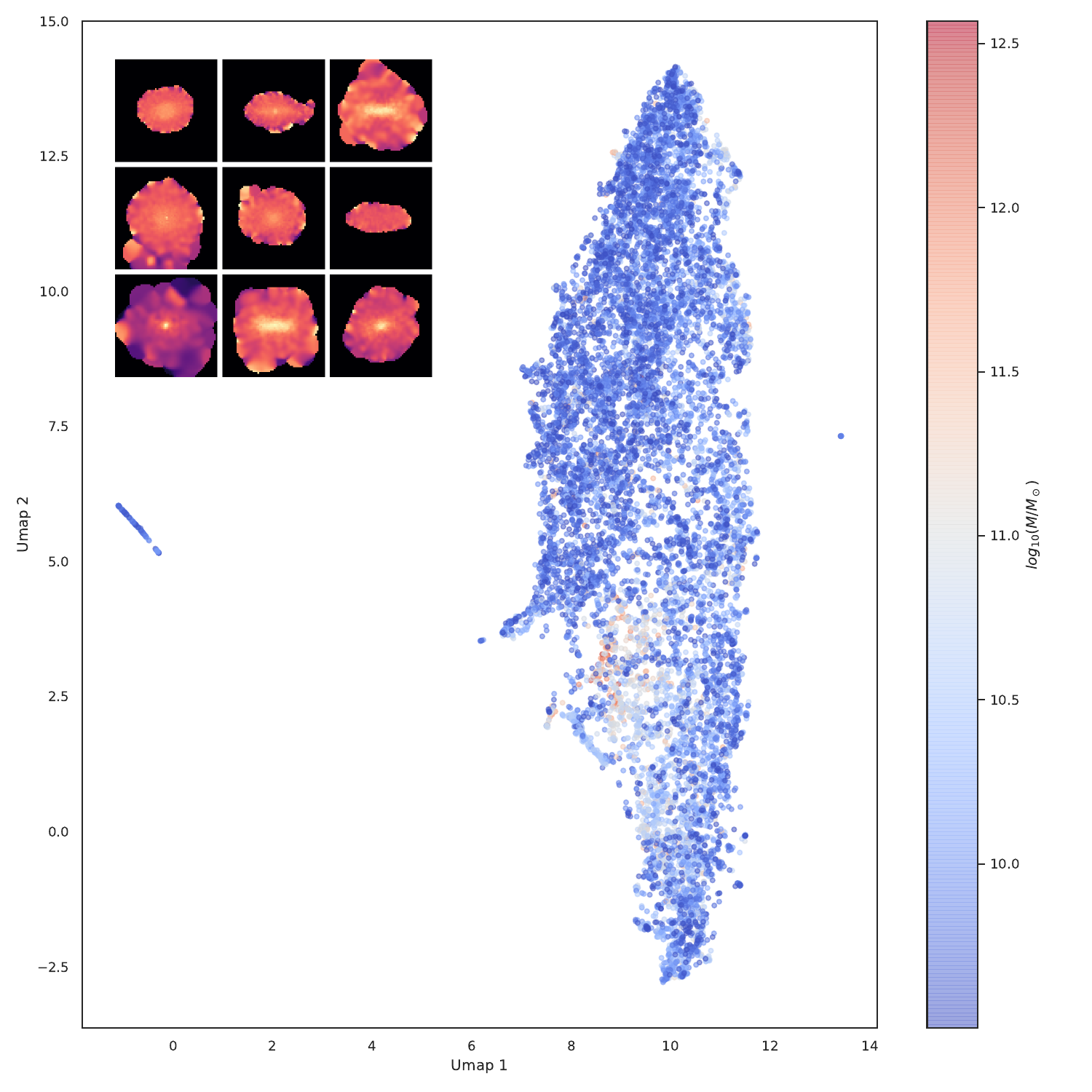}
    }
    \subfigure[]
   {
    \includegraphics[width = .48\textwidth]{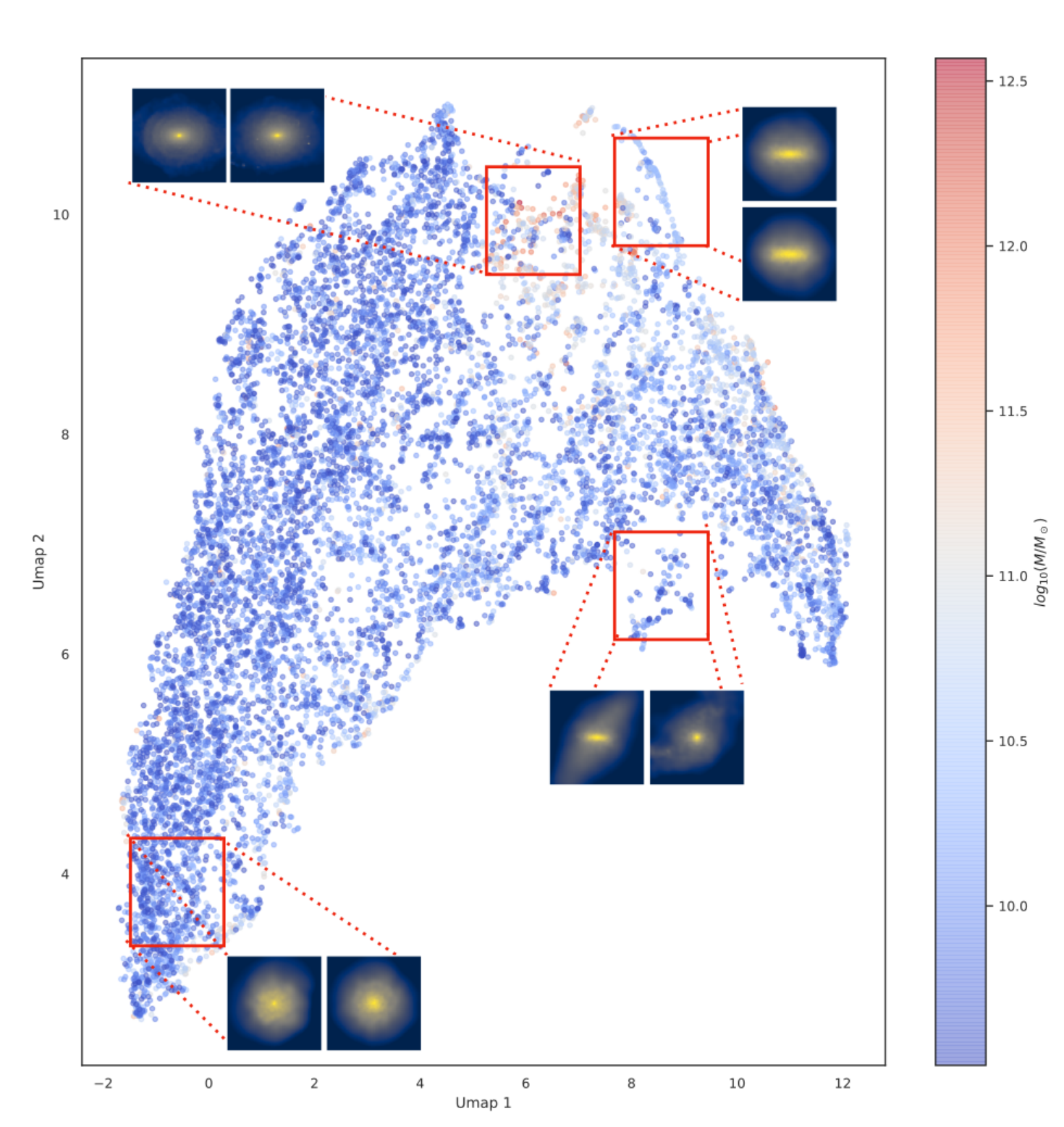}
   } 
    \caption{Outlier detection: UMAP projection with \textit{n\_components=2, n\_neighbors=5, min\_dist=0.001} calculated (a) on the PCA components of the 2D images to find outliers (b) after filtering outlier galaxy images by choosing galaxies with $\text{UMAP}_1<2$. We show some of the outliers at the top left of (a). The color map corresponds to the masses of the galaxies in $\log_{10} (M/M_\odot)$.
    An interactive version of the plot on the right can be accessed using an online dashboard referenced in the GitHub repository of this work (see Appendix \ref{sec:appendix_code_and_data})}
    \label{fig:umap projection 2d}
\end{figure*}

Figure \ref{fig:umap projection 2d} shows the UMAP embedding of the PCA scores of the two-dimensional galaxy images. UMAP separates a cluster of galaxies in the dataset that deviate significantly from the rest, as shown on the left in Figure \ref{fig:umap projection 2d}. These galaxies happen to have no stars in the outer regions and therefore are reconstructed using a significantly different linear combination of eigengalaxies. 

By employing a threshold on the UMAP coordinates we can filter out around $120$ galaxies. We rerun the analysis and fit UMAP on the PCA scores calculated on the filtered data. The resulting plot is shown on the right side of Figure \ref{fig:umap projection 2d}.
The lower left quadrant of Figure \ref{fig:umap projection 2d} mainly features spherical galaxies with little structure. In the top right section, galaxies with a prominent bar structure are aligned in a strip, whereas in the bottom right, galaxies characterized by a dominant diagonal feature form a separate cluster.
Interestingly, the upper middle region of Figure \ref{fig:umap projection 2d} is dominated by galaxies of higher mass.
An interactive version of this graph can be accessed using an online dashboard\footnote{URL: \url{https://github.com/ufuk-cakir/MEGS}}, where one can hover over each distinct point to see the corresponding galaxy image.

%\subsubsection{Reconstructing the joint distribution of stellar mass, metallicity and age from %broad band photometry}

\section{Summary \& Conclusions}
\label{sec: conclusion}
In this study, we have explored the application of Principal Component Analysis (PCA) to analyze galaxy images in both two-dimensional and three-dimensional cases, jointly modeling the mass, metallicity and age distributions. Our analysis has led to several findings and insights into the potential of using PCA to characterize galaxy morphology and perform morphological analysis.
   
%removed Fractional difference here
   \begin{enumerate}
      \item We have demonstrated that PCA can be effectively used to reconstruct galaxy images using a relatively small number of eigengalaxies (Fig. \ref{fig:one reconstruction sample galaxy 60 comp 2d}. The reconstruction accuracy was quantified using the reconstruction error (eq. \ref{eq:Reconstructin Error}), which measures the \U{difference} in pixel values between the original and PCA-reconstructed images. Our results (Fig. \ref{fig:rec_err_60_comp}) indicated that even with a modest number of 60 (215) eigengalaxies in 2D (\textbf{3D}), the PCA model achieved accurate reconstructions, with 90\% of the images having reconstruction errors below $0.022$ ($\mathbf{0.027}$). This points to the potential of PCA for efficiently representing complex galaxy morphology.
      \item We discussed the interpretability of the PCA model. Unlike more complex machine learning methods, PCA provides a straightforward interpretation of its components. Each eigengalaxy can be considered an image representing a specific morphological feature. Visualizing the top contributing eigengalaxies for different types of galaxies revealed their distinctive contributions in Fig. \ref{fig: decomposition_10} and showcased the potential of PCA for classifying galaxies based on their morphology.
      \item In terms of applications, we demonstrate the utility of PCA for performing morphological similarity searches in Fig. \ref{fig: nearest_neighbours}. By measuring Euclidean distances in the PCA eigenspace, we identified morphologically similar galaxies to a given sample galaxy. This suggests that PCA captures meaningful features of galaxy morphology, allowing for efficient similarity analysis and clustering.
      %\item RGB 
      \item We showcase outlier detection using UMAP as an application of our PCA model. By inspecting a UMAP projection of the PCA scores, we are able to filter out particular galaxies that do have no stars in the outskirts.
      \end{enumerate}
In conclusion, this study highlights the potential of PCA as a powerful tool for analyzing and characterizing galaxy morphology \U{and constructing a flexible, generative model for galaxy morphologies}. Its ability to efficiently represent images, interpret its components, and facilitate morphological similarity searches makes it a valuable approach in the field of astrophysics. \U{Our proposed PCA representation can be used to tackle the inverse problem of reconstructing physical parameters from observed data and in the case of 3D eigengalaxies it could even be used to tackle the inverse problem of deprojecting observed galaxies and reconstructing their physics parameters in 3D at the same time.} Future work could involve refining the PCA model, e.g. using non-linear PCA, exploring its applications in other astronomical datasets, and developing hybrid approaches that combine PCA with other machine learning techniques for even more comprehensive analyses of galaxy images.

\begin{acknowledgements}
      This project was made possible by funding from the Carl Zeiss Stiftung.
      %Part of this work was supported by the German
      %\emph{Deut\-sche For\-schungs\-ge\-mein\-schaft, DFG\/} project
      %number Ts~17/2--1.
\end{acknowledgements}


\bibliographystyle{aa}
\bibliography{bibtex/references}

\begin{appendix}
\section{Column-wise Mean of the Data Matrix}
\begin{figure}
    \centering
    \includegraphics[width=\hsize]{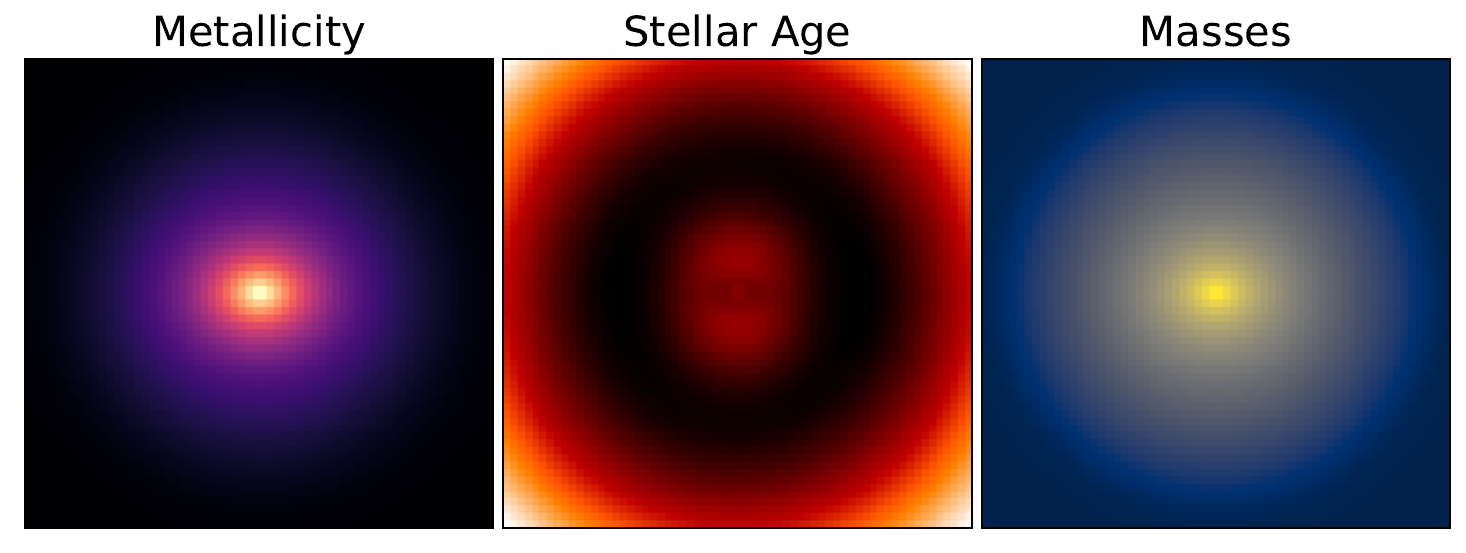}
    \caption{Columnwise mean of the Datamatrix as used in equation \ref{eq: centering the data}}
    \label{fig:mean galaxy images}
\end{figure}
\U{
The column-wise mean of the data matrix is computed within the PCA implementation of the \texttt{sklearn} library. This mean is visualized in Figure \ref{fig:mean galaxy images}.}

\section{Data Structure}
   \begin{figure}
   \sidecaption
   \includegraphics[width=\hsize]{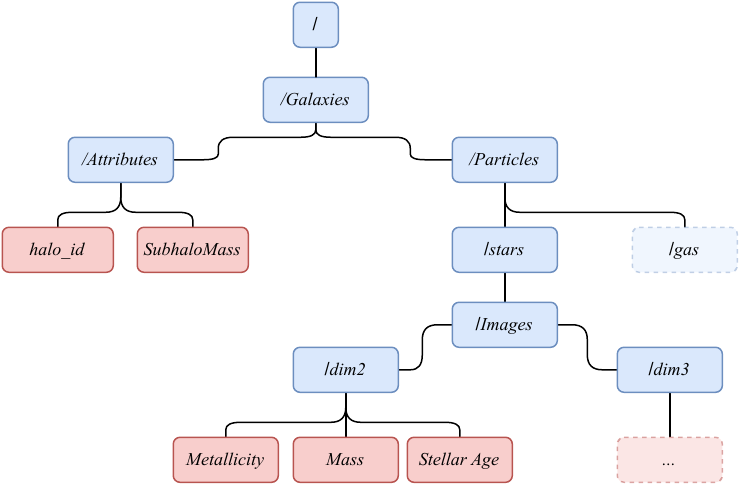}
      \caption{HDF5 File Structure, where each blue node represents a HDF5 group and red nodes are  datasets. The data file contains general galaxy parameters under the \textit{/Galaxies/Attributes} group such as the \textit{SubhaloID} in the simulation or the total mass of stars particles $M_\star$. The images of particles in the different fields (i.e \textit{metallicity}, \textit{mass}, \textit{stellar age}) are calculated in two and three dimensions and saved under the respective subgroup.
              }
         \label{fig: hdf5_file_struc}
   \end{figure}

\U{The dataset is stored as a HDF5 file, with the data structure shown in Figure \ref{fig: hdf5_file_struc}.
The dataset is structured into hierarchical groups and datasets. At the top level, general galaxy parameters such as the \textit{SubhaloID} from the simulation and the total stellar mass ($M_\star$) are stored within the \textit{/Galaxies/Attributes} group. These parameters provide essential metadata for each galaxy in the simulation. Additionally, the file contains images of particle properties across different fields, including \textit{metallicity}, \textit{mass}, and \textit{stellar age}. These images are calculated in both two and three dimensions and are saved under their respective subgroups. This structure allows for efficient access and storage of large-scale simulation data, facilitating analysis across multiple dimensions and properties.}

\section{Code and Data Availability}
\label{sec:appendix_code_and_data}
The source code for the morphology model developed in this study, named \texttt{MEGS}, is made available under an open source license and is publicly hosted on GitHub.\\
To facilitate a wider community's usage and contributions, we have ensured that the repository is well-documented. The repository includes comprehensive documentation hosted on \emph{ReadTheDocs} that provides an overview of the project, installation instructions, and a guide on how to use the software. 

The MEGS repository, which contains code, additional documentation, and interactive dashboards, can be found at the following URL: \url{https://github.com/ufuk-cakir/MEGS}

The cleaned data set with outliers removed is publicly available on Zenodo under the URL:
\url{https://zenodo.org/record/8375344}

 \begin{figure*}
   \centering
   \subfigure[Metallicity]{
      \includegraphics[width=.75\hsize]{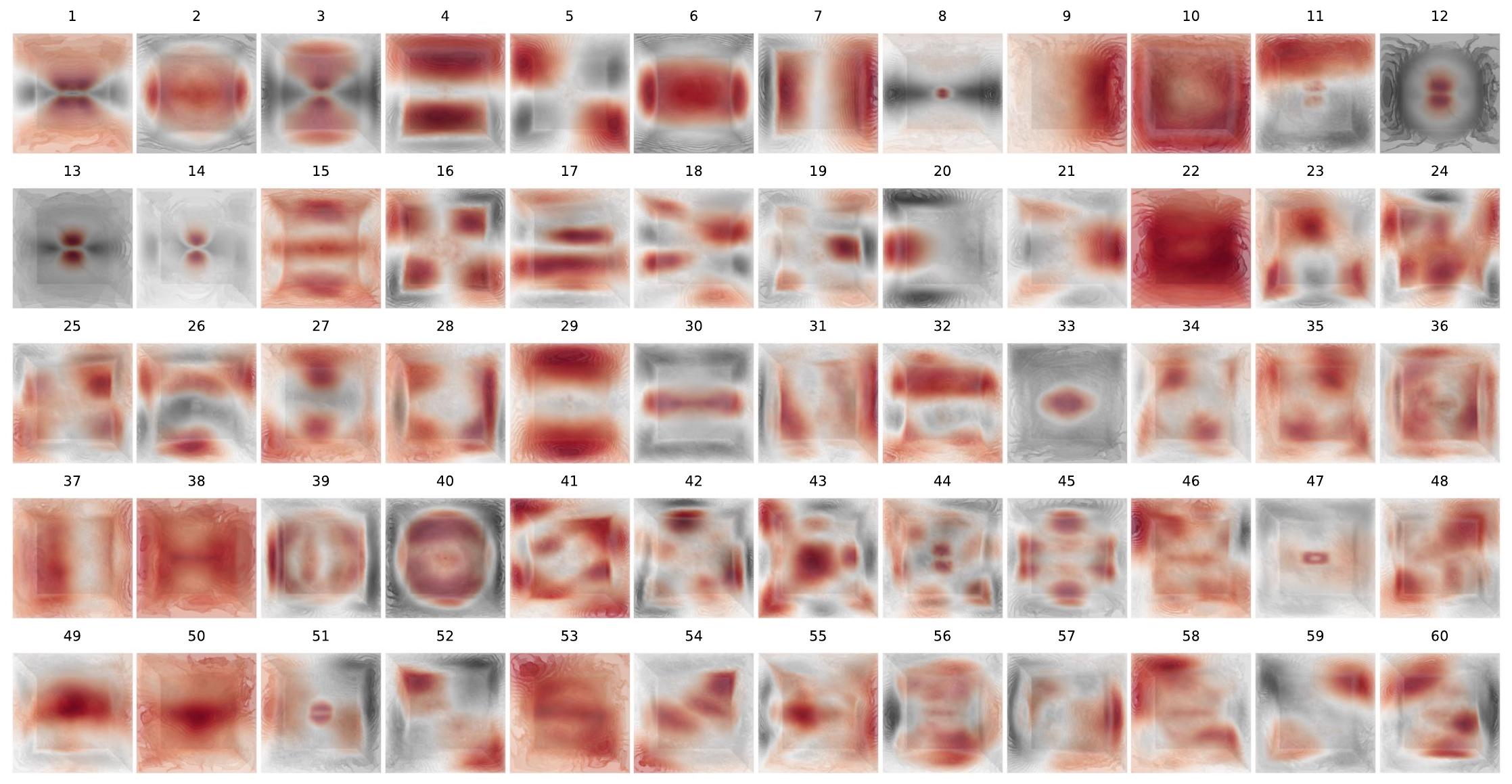}
   }
   \subfigure[Stellar Age]{
      \includegraphics[width=.75\hsize]{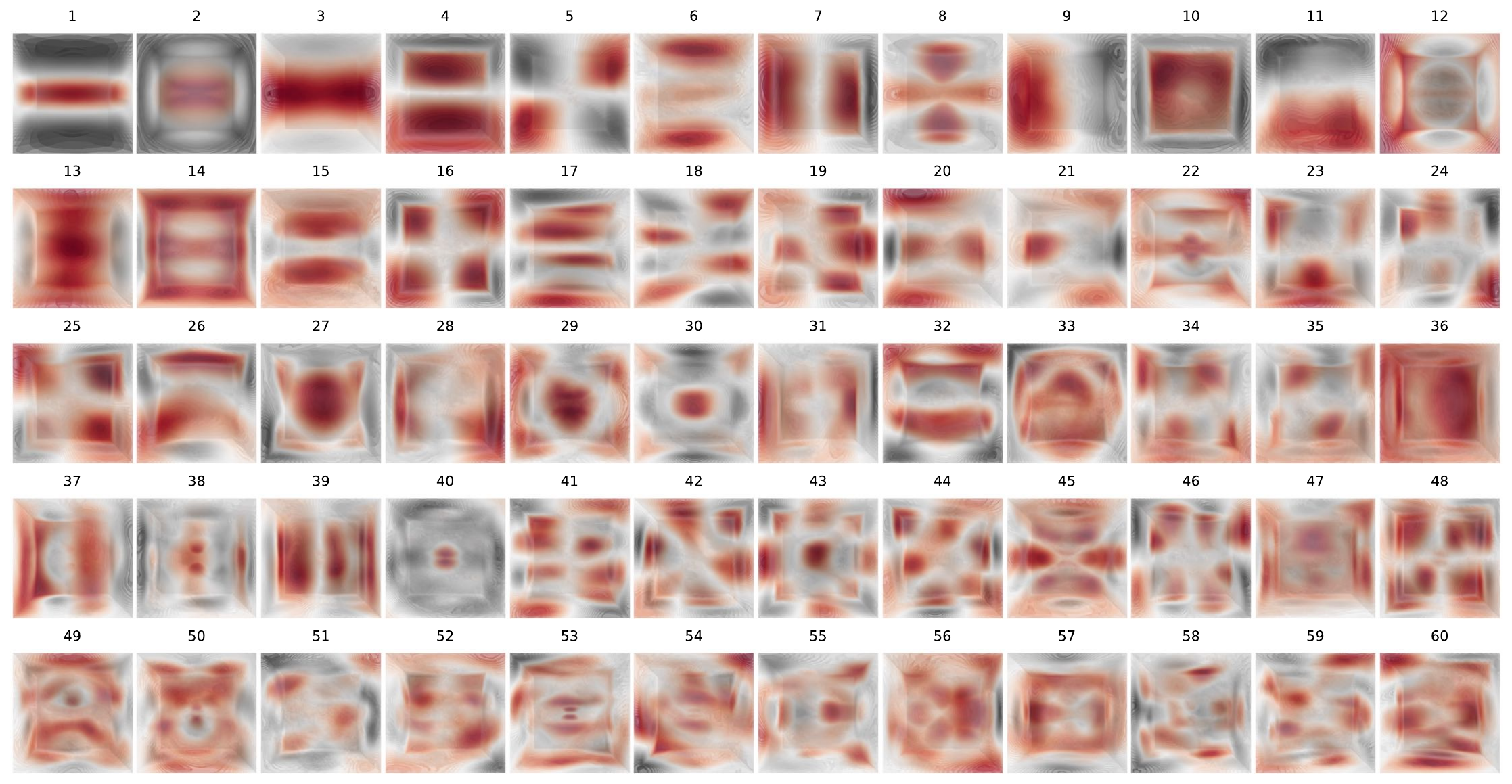}
   }
   \subfigure[Masses]{
     \includegraphics[width=.75\hsize]{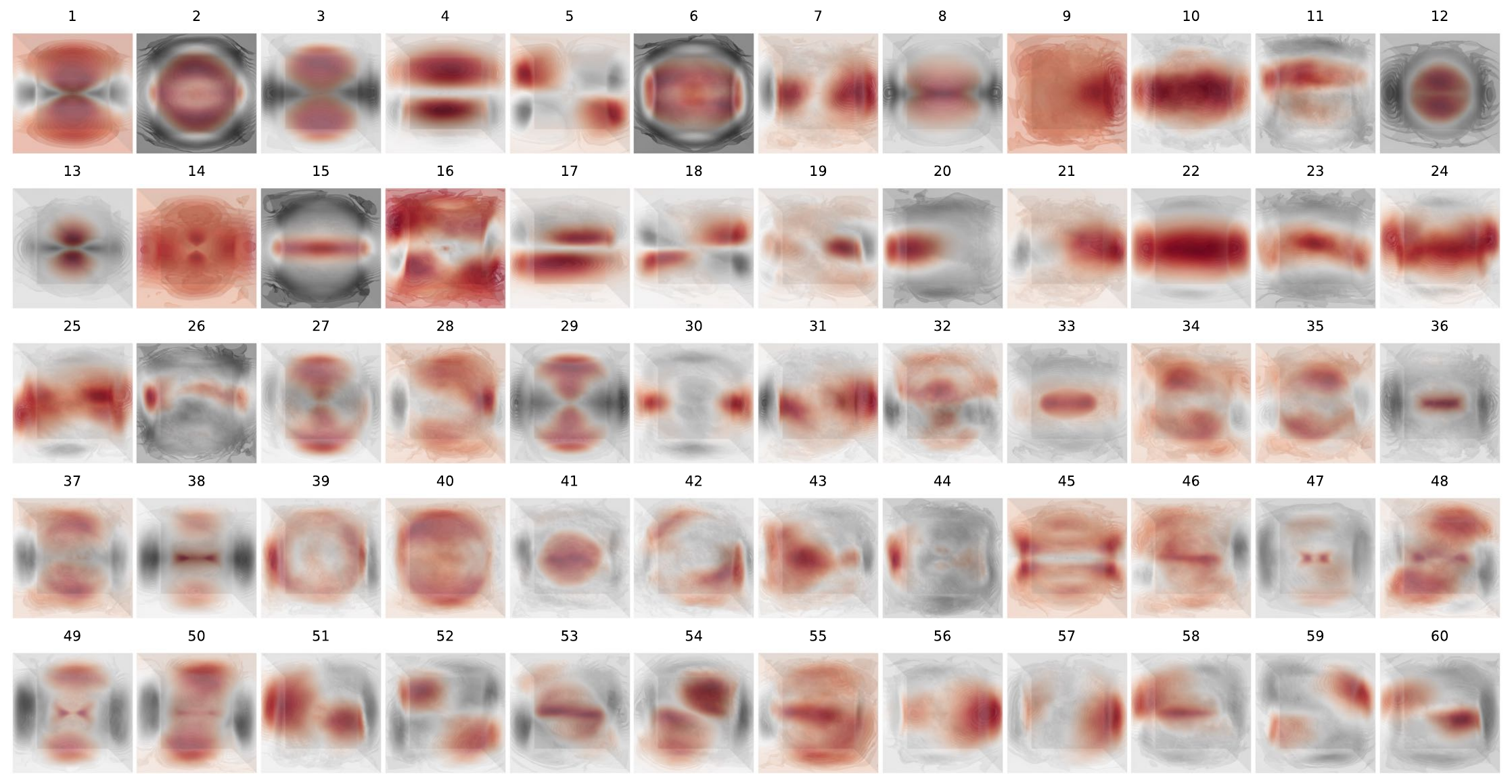}
   }
   \caption{First 60 Eigengalaxies from PCA in 3D: For better visualization we show the images in the x-z direction}   
   \label{fig: eigengalaxies 3D}
\end{figure*}

%\begin{figure}
%%    \centering
%    \includegraphics[width=\hsize]{figures/eigengalaxies/explained_variance_60_comp.png}
%    \caption{The plot illustrates the explained variance ratio resulting from 400 separate PCA calculations keeping 60 eigengalaxies. These calculations were performed on a subsample composed of 75\% randomly selected galaxies from the dataset. The histogram provides insights into the effectiveness of the PCA in capturing the underlying patterns and structures within the randomly subsampled data.}
%    \label{fig:400_evr_calc_60_components}
%\end{figure}

   \end{appendix}
\end{document}